# Understanding and Controlling V-Doping and S-Vacancy Behavior in Two-Dimensional Semiconductors – Toward Predictive Design

*Shreya Mathela[1‡], Zhuohang Yu[2‡], Zachary D. Ward[3], Nikalabh Dihingia[1], Alex Sredenschek[4], David Sanchez[2], Kyle T. Munson[2], Elizabeth Houser[2], Edgar Dimitrov[4], Arpit Jain[2], Danielle Reifsnyder Hickey[1,2,5,6], Humberto Terrones[3\*], Mauricio Terrones[1,2,4,5\*] and John B. Asbury[1,2\*]*

1. Department of Chemistry, The Pennsylvania State University, University Park, PA 16802, USA
2. Department of Materials Science and Engineering, The Pennsylvania State University, University Park, Pennsylvania 16802, United States
3. Department of Physics, Applied Physics and Astronomy, Rensselaer Polytechnic Institute, Troy, NY 12180, USA
4. Department of Physics and Astronomy, The Pennsylvania State University, University Park, Pennsylvania 16802, United States
5. Center for 2-Dimensional and Layered Materials, The Pennsylvania State University, University Park, PA 16802, USA
6. Materials Research Institute, The Pennsylvania State University, University Park, PA 16802, USA

†terroh@rpi.edu
#mut11@psu.edu
*jba11@psu.edu

‡ Denotes equal contribution

## ABSTRACT

Doping in transition metal dichalcogenide (TMD) monolayers provides a powerful method to precisely tailor their electronic, optical, and catalytic properties for advanced technological applications, including optoelectronics, catalysis, and quantum technologies. However, doping efficiency and outcomes in these materials are strongly influenced by the complex interactions between introduced dopants and intrinsic defects, particularly sulfur vacancies. This coupling between dopants and defects can lead to distinctly different behaviors depending on doping concentration, presenting significant challenges in the predictable and controlled design of TMD properties. For example, in this work we systematically varied the p-type vanadium (V) doping density in tungsten disulfide ($WS_2$) monolayers and observed a transition in doping behavior. At low concentrations, V-dopants enhance the native optical properties of $WS_2$, as evidenced by increased photoluminescence, without introducing new electronic states. However, at higher concentrations, V-dopants promote the formation of vanadium–sulfur vacancy complexes that generate mid-gap states, with energies that can be precisely tuned by controlling the vanadium concentration. Using a combination of excitation- and temperature-dependent photoluminescence microscopy, atomic-resolution scanning transmission electron microscopy, and first-principles calculations, we identify attractive interactions between p-type V-dopants and n-type monosulfur vacancies. Our results provide mechanistic understanding of how enthalpic dopant-defect interactions versus entropic effects govern the balance between property enhancement versus perturbation of transition metal dichalcogenides and suggest a pathway toward the rational design of doping strategies for next-generation optoelectronic, catalytic, and quantum devices.

## INTRODUCTION

Two-dimensional (2D) transition metal dichalcogenides (TMDs) such as $MoS_2$ and $WS_2$ offer exceptional potential for next-generation electronic, optical, and quantum technologies due to their unique optoelectronic properties.[1-3] Dopant engineering, the introduction of controlled amounts of foreign atoms, is a powerful strategy to tailor these properties. For instance, doping can tune bandgap energies[4], enhance photoluminescence[5], optimize carrier mobility and conductivity[6], modify strain[7], or introduce magnetic and spin-polarized states for advanced quantum information science applications.[8] However, the inherent confinement within the basal planes of monolayer TMDs often requires relatively high concentrations of substitutional dopants to tune free carrier densities due to the reduced screening caused by their lower dimensional morphology. This often leads to strong interactions of dopants with intrinsic defects, most notably monosulfur vacancies ($S_V$), which are particularly prevalent in these materials. These dopant-defect interactions depend on the dopant type (n-type or p-type) and concentration, often resulting in concentration-dependent changes in material properties that are difficult to predict. Such complex behavior makes it challenging to design synthesis and doping strategies for TMDs with electronic and optical properties[9-13] tailored for specific applications.

Prior work demonstrates that subtle variations in dopant concentration, local chemical environment, strain, and synthesis conditions can drastically alter how dopants influence material properties. For example, dilute rhenium doping has been shown to suppress $S_V$ formation in $MoS_2$, improving field effect transistor mobility and suppressing intrinsic defects.[14] However, other measurements of doping behavior predict that rhenium–$S_V$ complexes can trap carriers and negate the intended n-type behavior.[9,15] These seemingly contradictory findings reveal critical knowledge gaps in how dopant-defect interactions evolve across doping regimes, synthesis methods, and materials systems. Addressing these gaps requires approaches that integrate scalable TMD growth methods, understanding of concentration-dependent effects, and experimental validation coupled with computational description.[19]

In this work, we sought to address these knowledge gaps by systematically examining a range of dopant-defect interactions in vanadium (V)-doped $WS_2$ monolayers as they transition between two different doping behaviors: enhancement of photoluminescence (PL) in $WS_2$ at low doping levels to suppression of PL and introduction of mid-gap states[16] at higher doping levels. Importantly, we leveraged the p-type doping of vanadium in $WS_2$ to compare with prior studies of dopant-defect interactions. The asymmetric n-type intrinsic defect chemistry of TMDs makes the study of p-type dopants particularly useful. We used a combination of atmospheric-pressure chemical vapor deposition (APCVD) synthesis, excitation- and temperature-dependent PL spectroscopy/microscopy, atomic-resolution scanning transmission electron microscopy imaging (STEM), and first-principles calculations to examine the concentration-dependent evolution of V-dopant behavior in $WS_2$. We found that at low vanadium concentrations (<0.17 atom%), V-dopants suppress non-radiative Auger relaxation pathways and enhance the photoluminescence of $WS_2$ without significantly perturbing its electronic structure. However, at higher doping levels (≥1.1 atom%), vanadium–sulfur vacancy complexes (V-$S_V$) form that introduce mid-gap states that quench PL and have energies dependent on vanadium concentration. Atomic-scale STEM imaging confirmed the presence of V-$S_V$ dopant-defect complexes in V-doped $WS_2$.

Finally, we used first-principles computational methods to show that these mid-gap states originate from electronic interactions of V-dopants and sulfur vacancies as they form complexes, particularly in WS$_2$ monolayers with higher concentrations of vanadium. We identified an energetic stabilization of S$_V$ defects when they pair with vanadium atoms, which provides an enthalpic driving force to form such complexes during synthesis and doping of the material. However, at lower V-doping concentrations, dopants and S$_V$ defects have lower probabilities to form close complexes, which prevents the formation of mid-gap states. In this case, the p-type V-dopants scavenge electrons thermally emitted from sulfur vacancies, which leads to enhancement of the PL from the WS$_2$ monolayers by suppression of Auger relaxation processes. We reveal that the concentration-dependent V-doping behavior is governed by the interplay between the entropic driving force to randomly distribute dopants and S$_V$ defects throughout the material versus the attractive enthalpic interactions between p-type dopants and n-type defects that are present at higher dopant concentrations. Understanding and ultimately controlling this interplay of thermodynamic factors should provide a framework to predict where such transitions in doping behavior may occur, which will help design synthesis and doping methods for TMDs that are optimized for display and sensor technologies, electronic materials with high mobilities and reduced carrier scattering, or quantum information science with spin polarizable dopant and electronic states, among others.[17-26]

**RESULTS AND DISCUSSIONS**

We synthesized undoped and vanadium-doped WS$_2$ monolayers using an APCVD method previously described (**Figure 1a**).[8] **Figure S1** presents atomic-scale annular dark-field (ADF) STEM images of undoped WS$_2$, as well as for three different vanadium doping concentrations: 0.17%, 1.1%, and 3.4%, where the vanadium atoms are identified by their darker contrast in the images. Raman scattering spectra were collected on samples with varying vanadium concentrations using a 532 nm excitation laser, which lies ~0.33eV above the bandgap of monolayer WS$_2$ and enables near-resonant enhancement of defect-activated modes.[27-29] The intensity of the defect-assisted LA(M) phonon mode (~175 cm$^{-1}$) increases with rising vanadium concentration (**Figure S2**), indicating greater structural disorder due to the incorporation of V atoms into the WS$_2$ lattice.[28]

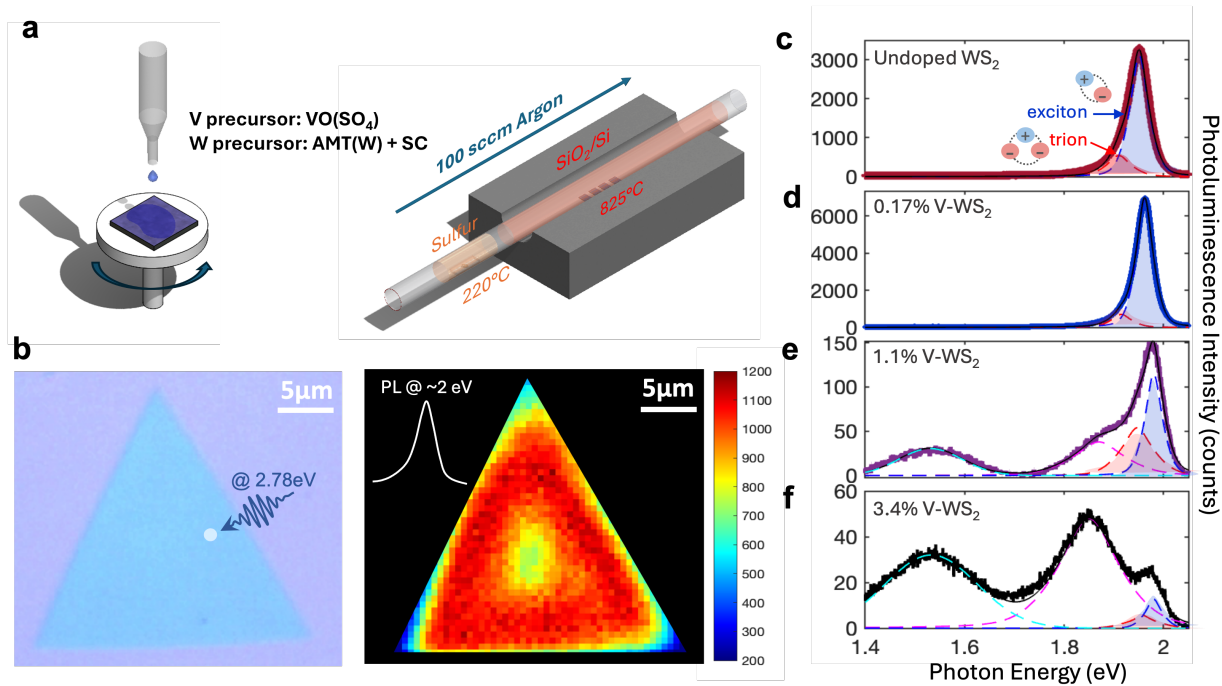

**Figure 1. Synthesis and room-temperature optical characterizations of ML V-WS$_2$:** (a) Schematic illustration of APCVD process for ML V-WS$_2$ films deposited on SiO$_2$-coated silicon substrate. (b)(left) Optical image of a flake that was excited by a 445 nm laser. The gray circle denotes the spatial distribution function. (right) PL map of the flake on the left centered at ~2 eV revealing the heterogeneity in the PL intensity across the flake. PL spectra of (c) undoped, (d) 0.17%, (e) 1.1%, and (f) 3.4% V-doped WS$_2$ films. The PL spectra were fit with two pseudo-Voigt curves to determine the contribution of A-excitons (blue) and trions (red) to the PL spectra. The peaks arising at 1.85 eV and 1.53 eV are fit with gaussian curves.

We examined the optical properties of V-doped WS$_2$ monolayers using spatially resolved PL imaging. Previous research demonstrated that emissive properties of TMDs can vary between crystallite edges and centers.[30, 31] This variation likely arises from a high density of structural defects such as Sv[31] at the growth edges and the presence of multilayers nucleating at the flake centers.[30] Consistent with these observations, spatially heterogeneous PL emission was observed in both pristine WS$_2$ (**Figure 1b**) and V-doped WS$_2$ monolayers (**Figure S3**), with the emission intensity at the edges and center regions being approximately half that of the intermediate regions (**Figure S4**). To minimize the influence of structural defects at the growth edges and multilayer regions, we focused our studies on the brighter intermediate areas of the WS$_2$ and V-doped WS$_2$ flakes, as indicated by the gray circle in **Figure 1b**. This gray circle also denotes the spatial distribution function on the same scale as the 5 μm scale bar.

PL emission spectra measured from the intermediate spatial regions of WS$_2$ monolayers with varying V-dopant concentrations are presented in **Figure 1c-f**. Under 445 nm excitation at 5.34 kW/cm$^2$ (2.00 µW average laser power), the PL spectra revealed a significant dependence on vanadium doping. We kept the integration times, excitation intensities, and optical geometries the same for all samples, allowing us to quantitatively compare the relative changes of their PL intensities. In pristine WS$_2$, the emission peak at ~1.97 eV could be decomposed into trion (X$^-$) and neutral exciton (X$^0$) components using pseudo-Voigt fitting (**Figure 1c**, red and blue shaded peaks). The pristine WS$_2$ sample is natively n-doped due to the combined effects of S$_V$ defects[32-35], hydrogenated defects on the SiO$_2$ substrate[36], and sodium atoms from the surfactant salt that are introduced during the APCVD growth.[37, 38] These n-type doping effects lead to the observed ratio of the trion-to-neutral exciton emission peaks of the pristine sample $I_{X^-}/I_{X^0}$ to be 0.2. This ratio corresponds to a free electron density of 5×10$^{11}$ cm$^{-2}$ based on a mass action model that was first proposed by Siviniant et al.[39, 40] The detailed calculation for the density of unbound electrons at the Fermi/defect level of WS$_2$ is provided in **note S1**.

Importantly, the incorporation of low vanadium concentrations (0.17 atom%) results in a significant enhancement of the A-exciton photoluminescence intensity[41] and a corresponding reduction in the trion-to-neutral exciton emission ratio (**Figure 1d**). Detailed fitting of the PL spectrum for the 0.17 atom% V-doped WS$_2$ monolayer revealed a decrease in the $I_{X^-}/I_{X^0}$ ratio to 0.1, which corresponds to a reduced free electron density of 2×10$^{11}$ cm$^{-2}$. This indicates that V-doping effectively compensates for the intrinsic n-type doping of WS$_2$, leading to lower electron densities and a PL spectrum dominated by radiative exciton recombination. The reduction in free electron density also mitigates nonradiative Auger recombination processes[42], resulting in enhanced PL from photogenerated excitons within the sample.

However, further increases in the V-doping density led to a marked reduction of PL emission intensity and the introduction of new mid-gap states that give rise to lower energy emission. For example, the sample with ~1.1 atom% concentration of vanadium dopants at tungsten sites exhibited a pronounced reduction of A-exciton emission and introduction of lower-energy emission bands around 1.85 eV and 1.53 eV (**Figure 1e**) even in the room-temperature PL spectrum. Further increase of the V-dopant concentration to 3.4 atom% led to a greater reduction of excitonic emission and the enhancement of the intensity of both new, lower-energy peaks

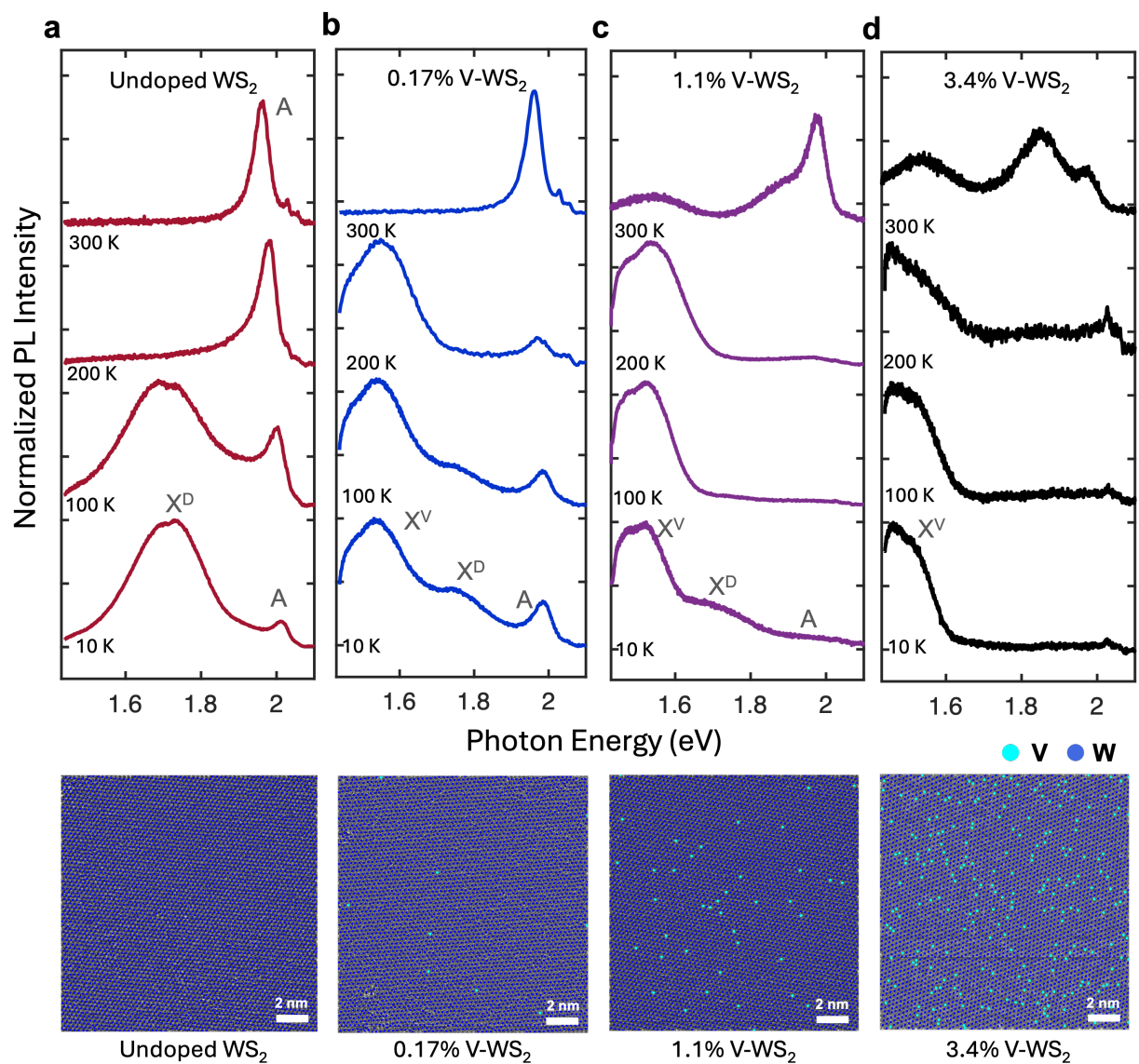

**Figure 2. Understanding the defect-related PL emission in V-WS$_2$:** Temperature-dependent PL spectra (top) and false-colored ADF-STEM images (bottom) from (a) undoped WS$_2$, (b) 0.17 at.% V-WS$_2$, (c) 1.1 at.% V-WS$_2$, and (d) 3.4 at.% V-WS$_2$. The low-temperature PL spectrum of undoped WS$_2$ exhibits a broad PL emission band at 1.65 eV that arises from defect-related excitons (X$^D$), while the PL spectrum of doped WS$_2$ reveals a new feature (X$^V$) arising because of interactions between the vanadium atom and sulfur vacancies. (Figure S1 shows raw STEM data)

(**Figure 1f**). These results reveal a concentration-dependent transition in V-doped WS$_2$ monolayers, where low dopant levels enhance the native optical and electronic properties by scavenging carriers introduced by defects and suppressing non-radiative pathways. In contrast, higher doping concentrations introduce lower energy emission features that indicate perturbation of the native electronic and structural properties of the WS$_2$ monolayers. We note that lower-energy PL features below the A-exciton optical gap have previously been observed in low-temperature spectra of TMDs and are typically attributed to defect-related excitons.[43-45]

To identify the microscopic origin of this concentration threshold at which the behavior of p-type vanadium dopants changes from enhancing native optical properties to perturbing them, we undertook a temperature-dependent photoluminescence study, which provides a sensitive probe of how exciton transport and recombination pathways evolve with temperature. **Figures 2a-d** represent temperature-dependent PL spectra of $WS_2$ samples doped with 0% (pristine), 0.17%, ~1.1%, and ~3.4% vanadium measured at temperatures between 300 and 10 K. A corresponding false-colored ADF-STEM image is displayed below each PL temperature series to display the V distribution (see **Figure S1** for raw STEM data). The spectra were measured using the same 445 nm excitation wavelength at 5.34 $kW/cm^2$ and the same integration times, excitation intensities, and optical geometries for all samples. The PL spectra measured at 300 K reveals the same variation of trion-to-neutral exciton emission ratio $I_{X^-}/I_{X^0}$ and low-energy PL emission features that appear in **Figure 1c-f**. Importantly, the PL spectra measured at lower temperatures enables us to suppress Auger relaxation processes resulting from free carriers emitted from dopant and defect states in the materials. We use this approach in comparison to room-temperature measurements to examine the defect-related excitonic states that are neutral and therefore more emissive in comparison to the same states that at higher temperature lead to free carrier formation and therefore undergo rapid Auger relaxation with corresponding lower radiative decay probabilities.

In pristine $WS_2$, the defect-related excitonic emission features are enhanced at lower temperatures, as compared to the A-excitonic and trion emission, as expected from prior literature precedent.[46] Additionally, the neutral A-exciton shifts to higher energy as the temperature is decreased[47], a well-known behavior resulting from lattice contraction[8] or due to lower density of states at the optical band-edge due to reduced thermal disorder and electronic state broadening.[48] Additionally, a lower-energy emission peak around 1.7 eV (~250 meV away from neutral A-exciton) appears below 150 K, which has been assigned to excitons bound to neutral $S_V$ defects. Chalcogen vacancies, whether unpassivated or filled, serve as active sites for the adsorption of gas molecules such as $O_2$, $H_2O$, or $N_2$. This molecular adsorption process, along with the presence of intrinsic defects created by the vacancies themselves, can lead to sub-gap PL, which we broadly refer to as defect-related exciton emission $X^D$.[49]

Furthermore, V-doped WS$_2$ monolayers exhibit an additional, more red-shifted PL feature at a lower energy than the X$^D$ emission, which is absent in the pristine sample. This new peak denoted peak X$^V$ emerges below ~1.7 eV, and its exact energy depends on the V concentration. In the 0.17% V-doped WS$_2$ monolayer sample (**Figure 2b**), the newly observed peak X$^V$ emerges around 1.53 eV, while the lower-energy X$^D$ emission peak appears around the same transition energy as the pristine sample. However, in the 1.1 and 3.4 atom% V-doped WS$_2$ monolayer samples, peak X$^V$ shifts to lower transition energies around 1.5 and 1.46 eV (**Figure 2c, d**), respectively. This pronounced red shift of X$^V$ with increasing vanadium concentration (**Figure S5**) indicates that the V-induced emissive state becomes progressively deeper in the bandgap as more V is introduced. To verify the reproducibility of trends observed in **Figure 2b-d**, we analyzed three sets of WS$_2$ samples synthesized on different days and doped with varying V concentrations. For each sample set, we measured temperature-dependent PL spectra of 5-7 monolayer flakes, fit the peaks to determine their energies, and calculated the standard deviations. These analyses allowed us to define the error bars presented in **Figure S6**, which reflect the uncertainty in the X$^V$ emission from the V-doped WS$_2$.

The evolution of the PL spectra with temperature underscores how vanadium doping modifies exciton recombination pathways. In 0.17% V-WS$_2$, the X$^V$ emission appears at 200 K and persists to 10 K, while the X$^D$ emission emerges only at 10 K. The X$^V$ state exhibits lower-energy emission than the X$^D$ and therefore should serve as the dominant emission site in the V-doped samples at lower temperature. The appearance of higher-energy emission from X$^D$ states at 10 K suggests that a fraction of excitons at this temperature were unable to migrate to X$^V$ sites within their excited state lifetime. These findings indicate that 1) X$^V$ states have relatively high ionization energies such that 200 K is sufficiently low to prevent them from ionizing and undergoing rapid Auger relaxation and 2) X$^V$ states are sufficiently sparse at 0.17 V% doping so that not all excitons are able to migrate to them within their excited state lifetime at 10 K. However, at elevated temperatures around 200 K, the increased exciton mobility allows them to reach lower-energy X$^V$ sites more efficiently, suppressing X$^D$ emission. Furthermore, as the vanadium concentration increased to 1.1 atom%, emission from X$^V$ states became even more dominant than X$^D$ emission at 10 K. This is consistent with the higher density of X$^V$ sites accessible to excitons in the more highly doped sample, which allowed excitons to diffuse to them despite their reduced mobility at 10 K. In the 3.4 atom% V-doped sample, the X$^V$ state emission peak is the sole feature observed at all temperatures below 200 K. This again is consistent with the increased density of $X^V$ sites in the more heavily doped sample, which was sufficient for excitons to reach them regardless of temperature.

The lower-energy PL emission of $X^V$ states suggests that they correspond to mid-gap states within the optical bandgap of V-doped $WS_2$. We note that no such mid-gap states were predicted by theoretical models of V-doped $WS_2$, where vanadium atoms replaced tungsten atoms in the lattice.[8] This suggests that vanadium alone cannot induce the lower-energy emission or the corresponding mid-gap states. This discrepancy strongly suggests that the experimentally observed mid-gap states arise from interactions between V-dopants and native defects such as sulfur vacancies. We conclude that as the V concentration increased among the V-doped samples, the formation of these V-$S_V$ defect complexes became more probable, resulting in the greater intensity of $X^V$ emission at higher dopant levels. Therefore, we assign the origin of $X^V$ states to the presence of dopant-defect complexes involving both V atoms and sulfur vacancies.

We used two approaches to verify the assignment that the new lower-energy PL emission observed in the V-doped $WS_2$ monolayers (**Figure 2**) arose from complexes of vanadium atoms and $S_V$ defects. First, we performed post-growth sulfur annealing in the 1.1 atom% V-doped $WS_2$ sample because this annealing step has previously been shown to heal $S_V$ defects.[50] **Figure 3a** compares PL spectra measured in the same $WS_2$ flake of the V-doped $WS_2$ sample at 77 K before versus after the sample was exposed to the sulfur annealing process. Following the annealing, both the sulfur vacancy defect peak $X^D$ and the vanadium-related defect emission peak $X^V$ were no longer detectable at 77 K, demonstrating that $X^D$ and $X^V$ excitonic emission were effectively eliminated through reduction of the sulfur vacancy density. We show that quenched $X^D$ and $X^V$ emission following post-growth annealing did not arise from modifications to $WS_2$ / substrate interactions, which could affect $X^D$ emission via charge transfer interactions (see **Note 2** in the Supplementary Information). We transferred V-doped samples onto chemically inert hexagonal boron nitride (hBN) substrates and annealed them in a sulfur environment in the same manner as for the sample depicted in **Figure 3a**. Similar to $SiO_2$, we observed the disappearance of $X^D$ and $X^V$ emission following post-growth annealing of V-$WS_2$ on hBN, indicating that sulfur vacancy healing rather than modifications to substrate-induced charge doping was responsible for elimination of both lower-energy emission features. These findings substantiate that the lower-energy defect-related excitonic emission peak $X^V$ in the V-doped $WS_2$ samples arose from the coupling of V-dopants with $S_V$ defects as discussed above and confirm that V alone did not introduce mid-gap states in $WS_2$.[21]

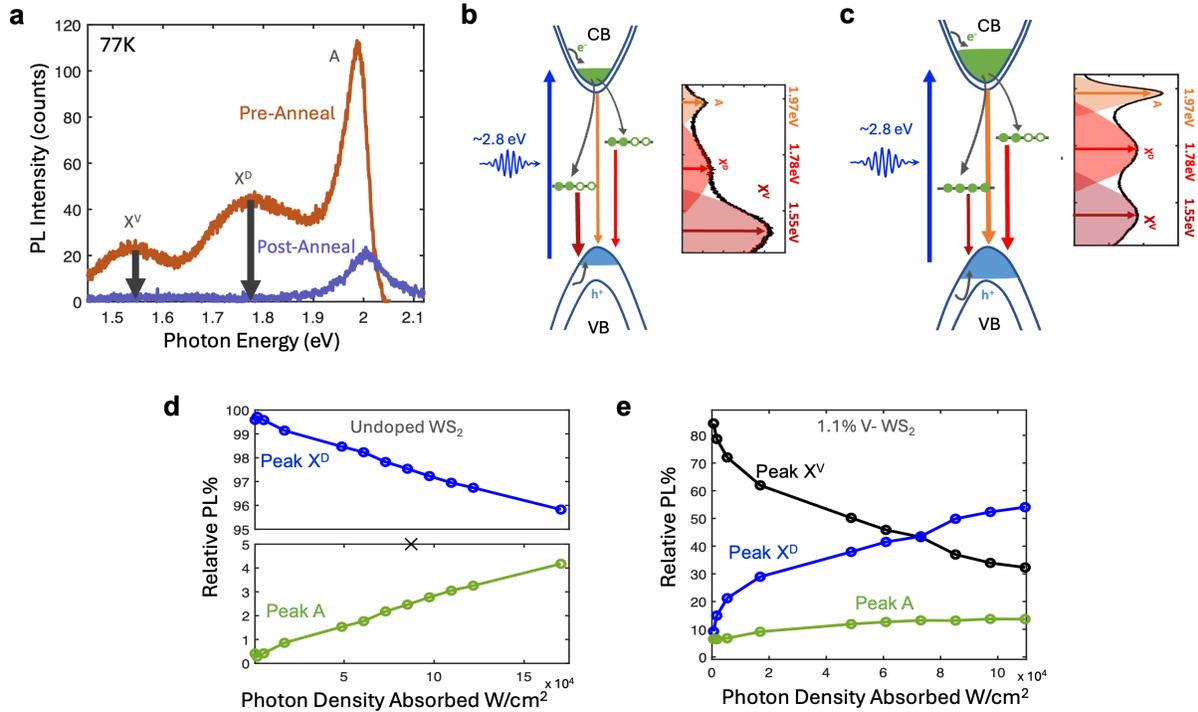

**Figure 3. S-annealing and excitation energy density dependent PL:** (a) PL spectra of 0.17% V-$WS_2$ taken at 77 K before and after S-annealing. After annealing, both the sulfur vacancy defect peak ($X^D$) and the vanadium-related defect emission peak ($X^V$) are no longer detectable, indicating that defect-related excitonic emissions were effectively eliminated through sulfur vacancy healing. (b,c) Schematic band diagrams showing that (b) at lower excitation intensities and low temperature, the PL spectra measured in 0.17% V-doped $WS_2$ samples are dominated by the distribution of defect states with a relatively small contribution from band-edge A-exciton emission, and (c) at higher excitation intensities, the limited density of mid-gap defect states can become saturated, leading to greater contributions of higher-energy defect states and band-edge A-excitons to the total emission from the sample. (d) Relative integrated photoluminescence (RIPL) intensities of peak $X^D$ and A plotted against the absorbed photon density (W/cm²) in a pristine sample, showing a slight decrease in the defect-related peak $X^D$ and an increase in A-excitonic emission with rising excitation intensity, consistent with sulfur vacancy defect saturation at higher intensities, and (e) 1.1% vanadium (V)-doped $WS_2$ monolayers displaying saturation of the defect peak $X^V$ at moderate excitation, while peak $X^D$ and A grow, indicating the impact of vanadium-sulfur vacancy pairing on emission behavior at elevated excitation levels.

Our second approach to verify that $X^V$ states arise from dopant-defect interactions was to use excitation intensity-dependent PL measurements to examine the density of the $X^V$ states. We hypothesized that if the $X^V$ state emission arises from the interaction of V-dopants and existing $S_V$ defects, then they should have limited densities of states because of the finite concentrations of both species in the $WS_2$ monolayer. To test this hypothesis, we compared the variation of the relative PL peak emission intensities of the A-exciton, $X^D$, and $X^V$ states as a function of excitation intensity, which is an established method to identify electronic transitions associated with defects. The method works because defect states can be saturated at moderate excitation intensities owing to their limited density of states.[51-55] The effect of excitation intensity on the distribution of PL emission amplitudes is illustrated in **Figures 3b and c.** At lower excitation intensities and low temperatures, the PL spectra measured in V-doped $WS_2$ samples were dominated by the distribution of defect states with a relatively small contribution from band-edge A-exciton emission. This is indicated in the inset of **Figure 3b,** showing a measured PL spectrum of a 0.17% V-doped $WS_2$ sample excited at 445 nm with 0.5 kW/cm$^2$ intensity at 10 K temperature. At higher excitation intensities, the limited density of mid-gap defect states can become saturated, which leads to increased contributions to the total emission of the sample from higher energy $X^D$ defect states and band-edge A-excitons.[51-55] This is indicated in the inset of **Figure 3c,** showing increased contributions from peak $X^D$ and the A-exciton emission peak to the measured PL spectrum of the same 1.1% V-doped $WS_2$ sample measured following 445 nm excitation with 110 kW/cm$^2$ intensity at 10 K temperature.[56]

To quantify the variation of the contributions of the various electronic states to the total emission of the pristine and V-doped $WS_2$ samples, we plot in **Figure 3d-e** the relative integrated PL (RIPL) intensity measured at 10 K of peak $X^V$, peak $X^D$, and A-exciton peaks versus excitation intensity absorbed by the sample (W/cm$^2$). In this analysis, the total integrated PL intensity across the entire spectrum for each excitation intensity is normalized to unity. Then, the PL spectra are fit to quantify the contribution of each transition in the spectrum. The RIPL intensity of each transition, normalized to the total emission intensity, is then plotted versus the corresponding excitation intensity used for that measurement. The data in **Figure 3d** for a pristine $WS_2$ monolayer contain only contributions from A-excitons and defect-related excitonic emission of peak $X^D$. The data reveal a modest decrease in the contribution from peak $X^D$ and an increase from A-excitonic emission consistent with a slight saturation of the sulfur vacancy defect states at higher excitation intensities.[51-55] This behavior is consistent with the relatively high sulfur vacancy defect density of $10^{11}$ to $10^{13}$ cm$^{-2}$ found in $WS_2$ monolayers synthesized by the APCVD method used here.[57-59]

In contrast, the data in **Figure 3e** were measured in the 1.1% V-doped $WS_2$ sample and exhibit emission from three peaks: $X^V$ states corresponding to complexes of vanadium atoms and $S_V$ defects, $X^D$ states arising from isolated sulfur vacancies, and A-excitonic states. At lower excitation intensities, the PL spectrum is dominated by $X^V$ state emission, which is consistent with the 10 K PL spectrum represented in **Figure 2b**. Importantly, the RIPL contribution of peak $X^V$ to the total PL emission from the sample saturates at moderate excitation intensities above 1 kW/cm$^2$. The relative contributions from peak $X^D$ and A-excitonic emission increase at these higher excitation intensities. This behavior is consistent with a limited density of vanadium atoms pairing with $S_V$ defects that give rise to the lower energy excitonic emission peak $X^V$ around 1.5 eV. The 3.4% V-doped WS$_2$ sample exhibited similar behavior (**Figure S7**). We note that the above experiments are fully reversible, indicating that the samples did not degrade under the highest excitation intensities examined here.

To verify the presence of dopant-defect complexes in the V-doped WS$_2$ monolayers, we used ADF-STEM imaging of samples with vanadium concentrations of 0.04%, 0.3% and 1.4%. **Figure 4a** depicts an ADF-STEM image of a 0.3% V-doped WS$_2$ sample. The intensity line profile in **Figure 4b** (left) reveals a dip in intensity at the tungsten and sulfur sites, which we attribute to a V-dopant adjacent to a S$_V$ defect. The green intensity line profile in **Figure 4b (right)** shows a dip at the tungsten site, corresponding to the V-dopant and no change in relative intensity at the sulfur sites, indicating the absence of a sulfur vacancy. **Figure S8** illustrates a Multislice simulated image of a V-doped WS$_2$ monolayer generated using the experimental parameters described in the Methods section. It shows a lower relative intensity at a S$_V$ defect adjacent to a vanadium atom, alongside tungsten and a site with two sulfur atoms, which is in good agreement with the line profile (**Figure 4b**) extracted from the experimental image (**Figure 4a**). We observed similar decreased intensities near random V-dopants in the 0.3% and 1.4% samples, which we attribute to S$_V$ defects coupled to V-dopants. **Figure S9** presents a larger-area image of 0.3% V-doped WS$_2$, showing several S$_V$ defects coupled to V-dopants. Additional S$_V$ defects were identified by extracting intensity line profiles, as shown in **Figure S10** and **S11**. In this analysis of the ADF-STEM images, we specifically designate as "S$_V$" any site that is missing a sulfur atom, without distinguishing between a single or double sulfur vacancy. This is because sulfur and vanadium have low atomic numbers (Z) and therefore low electron scattering probabilities. Consequently, the intensity line profiles extracted from the ADF-STEM images include the influence of background noise or non-Z-contrast signals present under the optimized detector conditions selected to best detect the low-Z atoms. Nonetheless, the data indicates the presence of dopant-defect complexes consistent with the assignment of the $X^V$ emission peak in the PL spectra. We note that such co-localization of V-dopants and $S_V$ defects was also reported by Zhang et al. in samples with higher vanadium concentrations.[8]

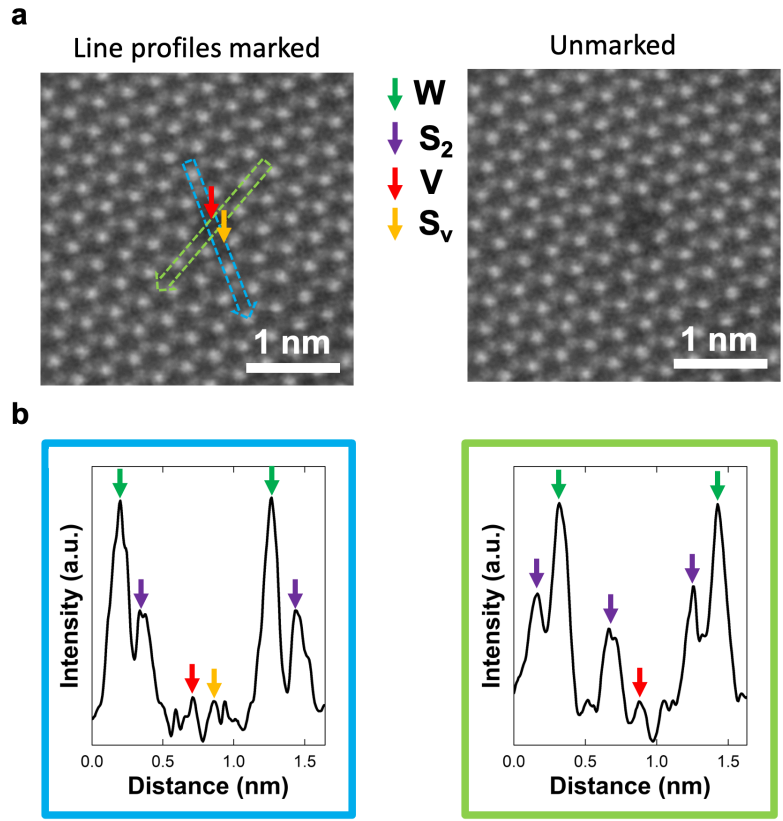

**Figure 4. Atomic-scale ADF-STEM images.** ADF-STEM image (a) marked and (b) unmarked versions of the 0.3% V-doped $WS_2$ showing V-$S_V$ site coupling. Intensity line profiles through the V site along directions that show the (c) $S_V$ site coupling (blue) and (d) no $S_V$ site (green).

Having established that V-$S_V$ complexes exist in V-doped $WS_2$ monolayers, we turned to first-principles computational methods to examine the origin of the concentration-dependent doping behavior of vanadium in $WS_2$. We employed Quantum Espresso ab initio computational methods to explore the overall electronic structure of V-dopants substituted at tungsten sites in monolayer $WS_2$ to understand how vanadium interacts with mono sulfur vacancy defects because these have higher densities than double vacancies. The in-plane lattice constants of $WS_2$ were set to a = b = 3.189 Å for both supercells used in this study with dimension 7 x 7 and 4 x 4. The separation between basal planes was set to c = 12.98 Å. The 7 x 7 supercell consisted of 146 atoms and was used to compute the energy of specific V-S$_V$ complexes with respect to the lowest-energy configuration, which was the dopant-defect complex with the smallest separation (**Figure 5a**), computed at the Γ point. The energies of other dopant-defect configurations above this value (Energy Above Hull) are also represented by color-coded circles corresponding to the highlighted locations in the inset of **Figure 5a**. The energies of excitonic transitions from these optimized nuclear configurations are represented in **Figure 5b** with the same color coding and were computed from the HSE calculated[60] band gaps. The splitting between the defect bands and the pristine WS$_2$ band edges was obtained from HSE density of states calculations as represented in **Figure 5c**. Additional computational details are provided in **Note 3** in the Supporting Information.

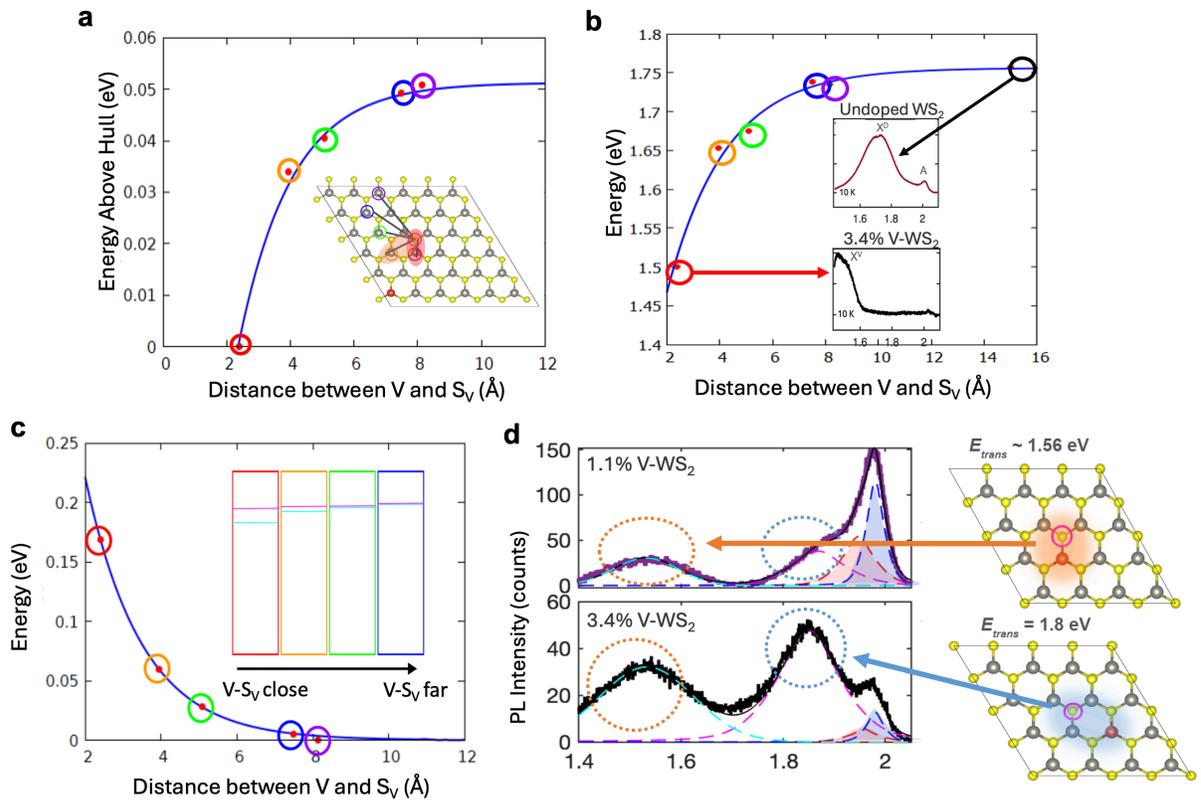

**Figure 5. Hybrid functional (HSE) calculations:** (a) The asymptotic growth of the energy above hull due to the coupling between the substitutional V atom and S$_V$ as a function of separation distance. The inset in (a) shows the V-S$_V$ paired WS$_2$, where the V (red) atom substitutes a W (gray) atom with corresponding substitutional V atoms circled in different colors and the S$_V$ defect circled in pink. (b) The evolution of the photon energy for paired substitutional V atom and S$_V$ with varying separation distances. The inset in (b) shows the PL spectra of undoped, and 3.4% V-doped WS$_2$ films. (c) The exponential decay of the splitting arising due to the coupling between the substitutional V and S$_V$ as a function of separation distance. The inset in (c) shows the evolution of the split defect bands where the border color represents the corresponding point in (a). (d) (left) Experimental PL spectra for V-doped WS$_2$ and (right) the two most stable V-S$_V$ defects in WS$_2$.

Remarkably, the variation of the computed transition energies for the various dopant-defect complex configurations compares quantitatively with the PL emission spectra measured in V-doped WS$_2$ samples at 10 K as indicated by the PL spectra appearing on the right-hand side of **Figure 5b** (reproduced from **Figure 2a** and **Figure 2d**). The computed excitonic transition energies of the dopant-defect complexes begin at the energy of ~1.5 eV of the X$^V$ peak in the PL spectra when the V-dopants and S$_V$ defects are in close proximity. Then, the computed excitonic transition energies asymptotically approach the X$^D$ peak energy of ~1.75 eV as the V-dopant/sulfur vacancy separation approaches 1 nm. At this separation, the V-dopants have little influence on the energetics of the S$_V$ defects. The blue curves in both figures represent exponential functions showing that the energies of the mid-gap states induced by V-dopant/S$_V$ defect interactions vary exponentially with their separation.

Furthermore, we used a 4 x 4 supercell consisting of 47 atoms to represent the optimized structure and energetics of various V-dopant/S$_V$ defect complex configurations as represented on the right-hand side of **Figure 5d** and **Figures S12 and S13**. These calculations were performed on a 2 x 2 x 1 k-point grid with spin-orbit coupling (SOC)[61] included and were used to show the formation of mid-gap states arising from the various dopant-defect complex configurations. These findings are consistent with prior DFT calculations of the electronic band structure of WS$_2$ revealing the formation of two localized mid-gap states associated with the presence of sulfur vacancies.[62-65] Our computational results indicate an energetic preference for pairing V-dopants and S$_V$ defects as adjacent neighbors rather than at separated distances, suggesting that there is an enthalpic driving force for the formation of these dopant-defect complexes in V-doped WS$_2$. Furthermore, DFT calculations of the electronic band structure (**Figures S12 and S13**) reveal the formation of two localized mid-gap states associated with the presence of sulfur vacancies[62-65] that are consistent with the lower-energy PL emission spectra measured in the V-doped WS$_2$ monolayer samples (**Figure 1**).

Combining the experimental evidence for the formation of V-dopant/S$_V$ defect complexes with the computational evidence for attractive interactions within these dopant-defect complexes suggests a framework with which to understand the transition in doping behavior as the vanadium concentration increased in WS$_2$ monolayers. In particular, the computational work suggests that at low dopant concentrations, the configurational entropy associated with randomly distributed V-dopants and $S_V$ defects dominates, enabling vanadium to enhance the native electrical and optical properties of the $WS_2$ host lattice by scavenging electrons that would be thermally emitted from sulfur vacancies at room temperature. However, as the V-dopant concentration increases, the probability increases that vanadium atoms and $S_V$ defects will be in close proximity. This allows the attractive interactions of V-dopants and $S_V$ defects to influence their pairing probability, which contributes to the formation of mid-gap states within the $WS_2$ band gap. Thus, at higher vanadium concentrations, the attraction of the p-type V-dopants with n-type $S_V$ defects introduces an enthalpic contribution that increases at higher V-doping levels and perturbs the electrical and optical properties of the V-doped $WS_2$ monolayers. The balance of entropic versus enthalpic thermodynamic factors revealed in this investigation provides an explanation for the transition in doping behavior of vanadium atoms in $WS_2$ monolayers and suggests that independent evaluation of these entropic versus enthalpic contributions will enable the predictive design of synthesis and doping strategies for TMDs with properties tailored for specific applications in next-generation optoelectronic, quantum, and catalytic technologies.

**CONCLUSION**

In summary, this study establishes a mechanistic framework for understanding how V-dopant–$S_V$ defect interactions evolve with doping concentration in TMDs. We used a combination of temperature- and intensity-dependent PL spectroscopy, ADF-STEM, and ab initio calculations to show a clear transition in V-doped $WS_2$ from a regime where low V-dopant concentrations enhance the native optical properties by suppressing non-radiative recombination, to a regime where higher vanadium concentrations promote the formation of V-$S_V$ complexes that introduce mid-gap states and modify the electronic structure of the host TMD. We show that this transition in V-doping behavior is governed by entropic effects that support random distributions of V-dopants and sulfur vacancy defects, allowing p-type V-dopants to scavenge electrons thermally emitted from sulfur vacancy defects. In contrast, at higher V-dopant concentrations, the attractive dopant-defect interactions become more important, increasing the probability that vanadium atoms and $S_V$ defects will pair with smaller separations. Our combined experimental and computational findings suggest that evaluation of dopant–defect interaction energies, relative to the entropy of randomly distributed dopants and defects, may offer a predictive framework for identifying concentration thresholds where the behavior of substitutional dopants such as vanadium change. The ability to predict such concentration thresholds will help support the predictive design of synthesis and doping strategies for the rational design of TMD-based materials for applications in optoelectronics, quantum photonics, spintronics, and catalysis.

## METHODS

### APCVD Growth:

WS$_2$ and V-WS$_2$ monolayers with different concentrations of V were synthesized via liquid-phase precursor-assisted ambient pressure chemical vapor deposition described elsewhere.[28, 29] We used ammonium metatungstate ((NH$_4$)$_6$H$_2$W$_{12}$O$_{40}$·xH$_2$O, AMT) and vanadium oxide sulfate (VO[SO$_4$]) as the tungsten and vanadium cation sources, respectively. Sodium cholate (C$_{24}$H$_{39}$NaO$_5$·xH$_2$O) functioned as a surfactant salt. We dissolved these precursors in deionized water and spin-coated the resultant solutions onto oxidized silicon substrates. The growth occurred in a two-zone furnace on Si/SiO$_2$ substrates with a 300 nm oxide layer. Metal precursors were made by mixing W-based solution with V-based solution, with the W-based solution prepared at a 1:4 volume ratio of ammonium metatungstate ((NH$_4$)$_6$H$_2$W$_{12}$O$_{40}$) to sodium cholate hydrate (C$_{24}$H$_{39}$NaO$_5$ • xH$_2$O) in deionized water. To achieve various doping levels, different molar concentrations of V-based solution were formulated using VO(SO$_4$) in deionized water. After spin-coating the combined solutions onto Si/SiO$_2$ substrates, the substrates were positioned within a quartz tube at the furnace's center. Simultaneously, 300 mg of sulfur powder was placed in an alumina boat in the upstream region in another heating zone. The reaction occurred with the substrate temperature reaching 825 ºC for a 15-minute growth period, during which sulfur was evaporated at 220 ºC. The process employed 100 sccm of Argon (Ar) as a carrier gas. Following synthesis, the furnace underwent cooling in an Ar atmosphere.

### Sulfurization:

Sulfurization experiments were carried out in a hot wall tube furnace with a base pressure of 20 mTorr. Samples were placed in an alumina crucible at the furnace center. The sample was heated to 700°C under a flow of 30 sccm ultra-high-purity argon as the carrier gas. Once the furnace reached 700°C, 15 sccm of H$_2$S gas was introduced into the furnace for an anneal duration of 60 min. After this reaction period, the H₂S flow was turned to 0 sccm, and the sample was cooled under 30 sccm of Ar.

**Transfer of V-WS₂:**

The PMMA-assisted wet-transfer method was employed to transfer the samples onto the desired substrate.[66, 67] A layer of PMMA (495 A6) was spin-coated onto WS₂ flakes on SiO₂/Si at 4000 rpm for 45 seconds, followed by baking at 180°C for 90 seconds. The samples were then immersed in a 1 M NaOH solution to etch the SiO₂. After the etching process, the PMMA film was thoroughly rinsed in deionized water for one hour before being transferred onto the target substrate (hBN/SiO₂, gold Quantifoil TEM grid).

**PL characterization:**

Photoluminescence measurements in a vacuum (approximately $10^{-6}$ Torr) were conducted using a microscope integrated with a closed-cycle helium cryostat (Montana Instruments, s100). The microscope employed a 100× 0.75 NA objective to focus a continuous-wave laser's output (Oxxius LBX-445, $\lambda_{ex}$ = 445 nm: 10 µW) onto the sample. After optical excitation, photoluminescence was gathered by the same objective, with stray laser light being eliminated through a 550 nm dichroic mirror (Thorlabs) and a 600 nm long-pass filter (Thorlabs). The collected photoluminescence was then directed into an optical fiber and concentrated onto the slits of a spectrograph (Princeton Instruments HRS-300SS, grating 300 grooves/mm) before detection by a back-illuminated CCD (Princeton Instruments, PIXIS-400BR). For temperature-dependent assessments, the sample underwent cooling from approximately 300 K to 10 K using the Montana Instruments s100 cryostat.

**Raman Spectroscopy:**

The Raman measurements were conducted using a Horiba LabRam HR Evolution confocal Raman system with a 532 nm DPSS laser and holographic gratings (1800 lines/mm). The system provides a spectral resolution of 0.5 cm⁻¹. A 100× objective lens was employed to focus the excitation laser, which had an approximate power of 1 mW. The typical acquisition time was 10 seconds, with an accumulation of 2 scans.

**STEM:**

ADF-STEM images were acquired using an FEI Titan G2 S/TEM microscope operated at 80 kV, with a probe convergence angle of 25.2 mrad and a beam current of 45-50 pA. To enhance the visibility of sulfur atoms and balance the contrast between tungsten (W) and sulfur (S), an ADF detector was used with collection angles of either 42-244 or 26-149 mrad. To extract intensity line profiles and reduce shot noise, slight Gaussian blur filtering (sigma = 2) was applied to certain images using ImageJ software. ADF-STEM images of vanadium-doped monolayer $WS_2$ (Figure S6) were simulated using the Multislice method with the PRISM algorithm, as implemented in the abTEM code.[68] The simulations were conducted under experimental-like conditions, with an accelerating voltage of 80 kV, eight frozen phonons, a convergence angle of 25.2 mrad, collection angles of 26–149 mrad, defocus of 25 Å, and slight aberrations ($C_{s(3)}$ = -0.0005 mm, $C_{s(5)}$ = 0.5 mm). For false-colored images, atom fitting of "W" and "V" atoms was performed using STEM images processed in ImageJ, followed by a custom-built Python script. First, the STEM images were filtered using bandpass filters with ranges of (3, 20) and (3, 40) to reduce noise from polymer residue and suppress scanning artifacts. Next, a Difference of Gaussian (DoG) filter was applied to the processed images to detect the atomic positions of "W" and "V" atoms. These positions were further refined using 2D Gaussian fitting, as implemented in the Atomap code.[69] To classify the atoms, we averaged the pixel intensities within a five-pixel region around each detected position. Atoms with lower average intensity were classified as "V" (colored cyan), while those with higher intensity were classified as "W" (colored blue). All atom assignments were visually checked against the raw data to confirm composition assignments. Several "W" atoms that were missed by the code but could be confidently identified as "W" based on visual inspection of the raw images in Figure S1 were manually added.

**DFT:**

The relaxed structures are obtained using the generalized gradient approximation (GGA) by Perdew, Burke, and Ernzehof (PBE)[70], while the HSE06 correction[60] is used to obtain the bandgaps from the density of states (DOS). The kinetic energy cutoff is set to 680 eV and the structures are relaxed until the total forces are smaller than 0.01 eV/Å and an scf convergence of $10^{-6}$ eV is achieved. Best fit equations are obtained using functions of the form, $E(d) = a +$

$b\exp(-cd)$ where $a$, $b$, and $c$ are best fit constants and $d$ is the distance between the V and $S_V$ defects. The values for the transition energy ($E_{trans}$) are obtained by proportionally shifting the B calculated excitonic transition energy obtained with SOC[61] to the experimental B exciton value (2.40 eV, $E_{B,\,EXP}$)[71] such that $E_{B,\,EXP}/E_{B,\,SOC} = E_{trans}/E_{blue,\,SOC}$, where $E_{blue,\,SOC}$ is the transition energy between the spin-up (blue) bands in the valence and lowest defect band at the K point. The wavefunction overlap is obtained by taking the product of the wavefunctions between the two energy levels responsible for the transition.

## AUTHOR INFORMATION

### Corresponding Author

* terroh@rpi.edu, mut11@psu.edu, jba11@psu.edu

### Author Contributions

The manuscript was written through the contributions of all authors. All authors have approved the final version of the manuscript. SM and ZY contributed equally to this manuscript.

## ACKNOWLEDGMENT

The authors SM and JBA are grateful for support from the Solar Photochemistry program within the Division of Chemistry, Geosciences and Biosciences, Office of Basic Energy Sciences, Office of Science within the U.S. DOE through Grant DE-SC0019349. MT thanks the Air Force Office of Scientific Research (AFOSR) through grant No. FA9550-18-1-0072. ND and DRH acknowledge generous support through startup funds from the Penn State Eberly College of Science, Department of Chemistry, College of Earth and Mineral Sciences, Department of Materials Science and Engineering, and Materials Research Institute. ND and DRH also acknowledge the Penn State NSF-MIP Two-Dimensional Crystal Consortium award (DMR-2039351) and the use of the Penn State Materials Characterization Lab. ZY and MT thank NSF for support under NSF DMR-2011839 through the Penn State MRSEC - Center for Nanoscale Science.

# Supporting Information

# Understanding and Controlling V-Doping and S-Vacancy Behavior in Two-Dimensional Semiconductors – Toward Predictive Design

*Shreya Mathela[1‡], Zhuohang Yu[2‡], Zachary D. Ward[3], Nikalabh Dihingia[1], Alex Sredenschek[4], David Sanchez[2], Kyle T. Munson[2], Elizabeth Houser[2], Edgar Dimitrov[4], Arpit Jain[2], Danielle Reifsnyder Hickey[1,2,5,6], Humberto Terrones[3*], Mauricio Terrones[1,2,4,5*] and John B. Asbury[1,2*]*

1. Department of Chemistry, The Pennsylvania State University, University Park, PA 16802, USA
2. Department of Materials Science and Engineering, The Pennsylvania State University, University Park, Pennsylvania 16802, United States
3. Department of Physics, Applied Physics and Astronomy, Rensselaer Polytechnic Institute, Troy, NY 12180, USA
4. Department of Physics and Astronomy, The Pennsylvania State University, University Park, Pennsylvania 16802, United States
5. Center for 2-Dimensional and Layered Materials, The Pennsylvania State University, University Park, PA 16802, USA
6. Materials Research Institute, The Pennsylvania State University, University Park, PA 16802, USA

†terroh@rpi.edu
#mut11@psu.edu
*jba11@psu.edu

‡ Denotes equal contribution

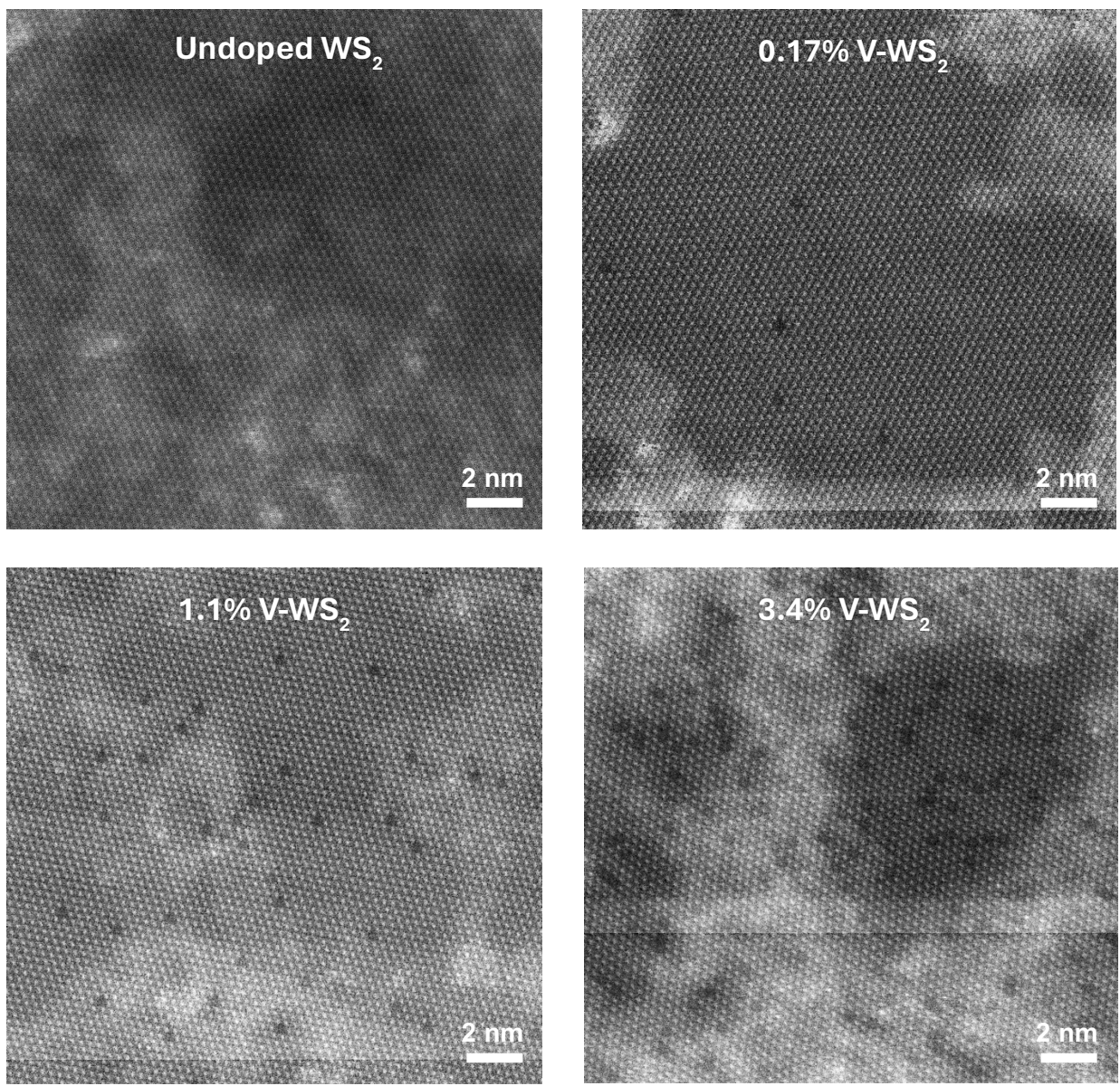

**Figure S1.** Annular Dark Field – Scanning Transmission Electron Microscopy (ADF-STEM) images of 0.17%, 1.1%, and 3.4% V-doped WS2.

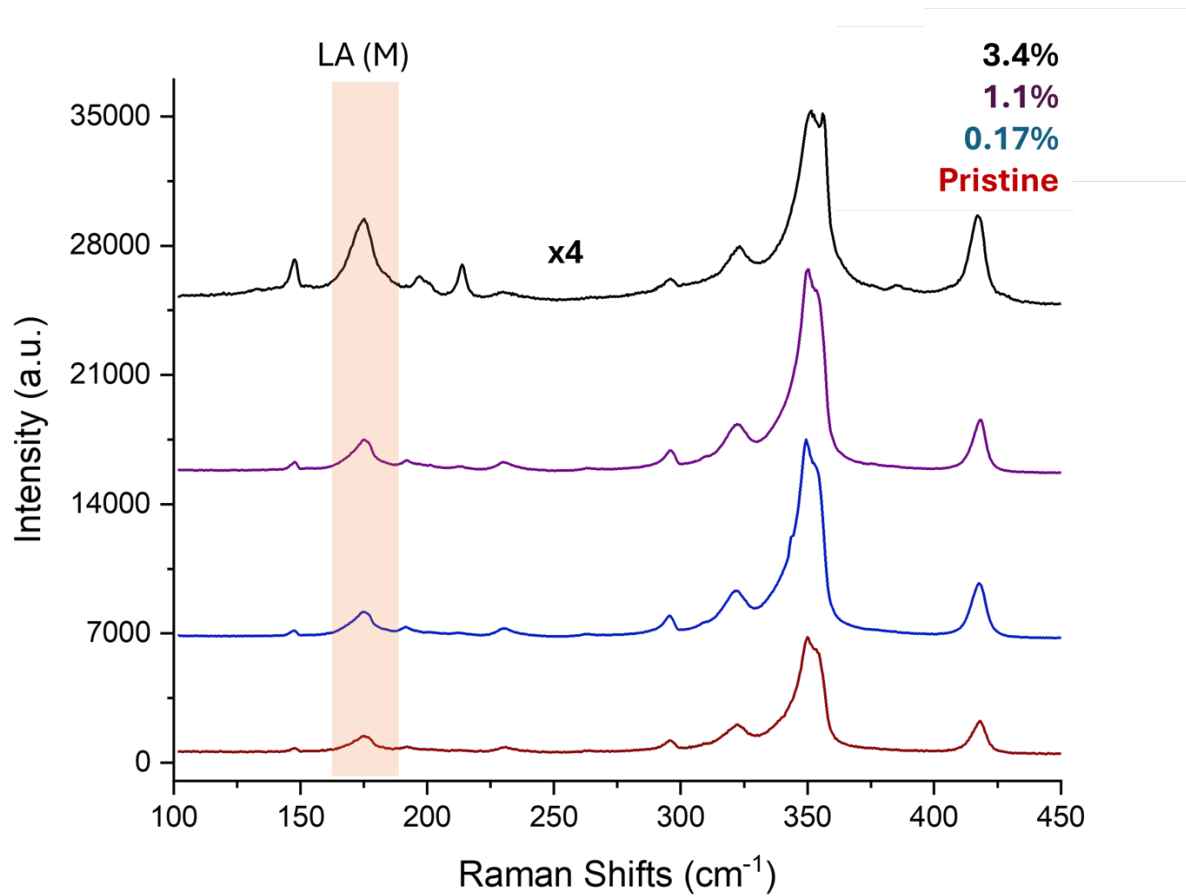

**Figure S2.** Raman spectra excited by 532 nm of pristine, 0.17%, 1.1%, and 3.4% V-doped $WS_2$. Defect-activated mode is highlighted in red region.

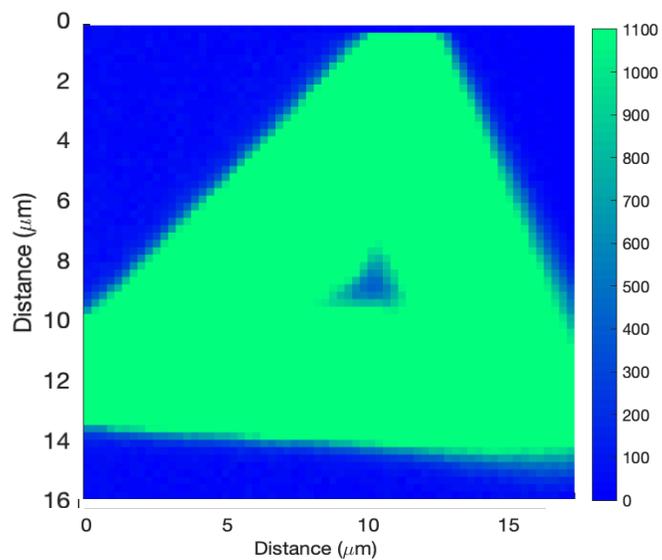

**Figure S3.** PL map of a 0.2% V-WS$_2$ flake centered at ~2 eV, showing variations in PL intensity across the flake.

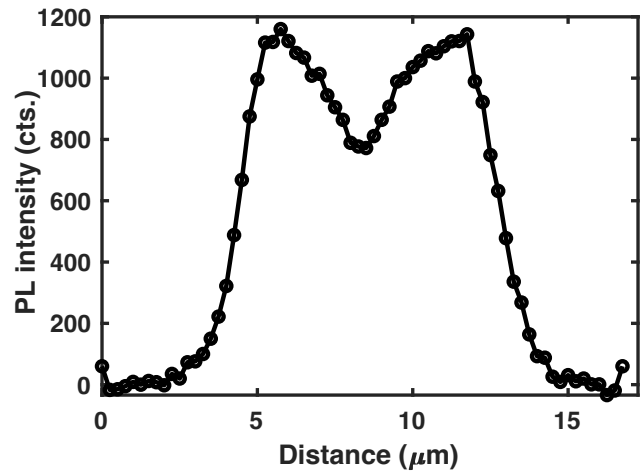

**Figure S4.** Intensity profile obtained from a linescan across the flake shown in **Figure 1b**, taken along a central axis from the left to right edge. The plot illustrates the spatial variation in signal intensity across the flake width.

**Note 1.** Calculation of the density of unbound electrons at the Fermi/defect level using the mass action law which describes the relationship between neutral excitons, trions, and excess electrons:

$$\frac{n_{X_o} n_{e-}}{n_{X-}} = \frac{4 m_{eff} K_B T}{\pi \hbar^2} \cdot \exp\left(-\frac{E_{X-}^b}{K_B T}\right) \quad \text{(Equation 1)}$$

where, $n_{X_o}$, $n_{e-}$ and $n_{X-}$ are the concentrations of excitons, unbound electrons, and trions at the Fermi level of monolayer WS$_2$, respectively. $m_{eff}$ is the reduced effective mass of monolayer WS$_2$, T is temperature, $K_B$ is the Boltzmann constant, $\hbar$ is the reduced Plank's constant, and $E_{X-}^b$ is the trion binding energy (~41 meV).

The effective mass of monolayer WS$_2$ is defined by $m_{eff} = \frac{m_{X_o} m_e^*}{m_{X-}}$

where, $m_{X_o} = \frac{m_e^* m_h^*}{m_e^* + m_h^*}$, $m_{X-} = \frac{m_e^*(m_e^* + m_h^*)}{(2 m_e^* + m_h^*)}$, $m_e = 9.109 \times 10^{-31}\ kg$

Here $m_{X_o}$ is the exciton mass, $m_{X-}$ is the trion mass, $m_e^*$ is the effective electron mass (0.31 m$_e$), and $m_h^*$ is the effective hole mass (0.42 m$_e$).

The concentration of trion/exciton ratio is directly evaluated from the area of trion and exciton peak spectra in the PL spectra[1,2] where $n_{X-}/n_{X_o}$ was calculated to be ~0.2217. $A_{trion}$ and $A_{exciton}$ are the spectral areas of excitons and trions in the PL spectra, respectively from the intensity ratio of the trion and A-exciton PL curves.

$$\frac{n_{X-}}{n_{X_o}} \sim \frac{A_{trion}}{A_{exciton}} \sim 0.2217$$

$$E_{X-}^b = E_{X_o} - E_{X-} \sim 0.0411\ \text{eV}$$

$E_{X-}^b$ and $E_{X-}$ are the peak positions of excitons and trions in the PL spectrum, respectively.

From Equation (1), we find that,

$$n_{e-} = 1.097 \times 10^{13} \exp(-38.75\ E_{X-}^b) \times \left(\frac{A_{trion}}{A_{exciton}}\right)$$

Inserting all parameters given above for 2.00 µW average laser power we obtain

$$n_{e-} = 5 \times 10^{11}\ cm^{-2}$$

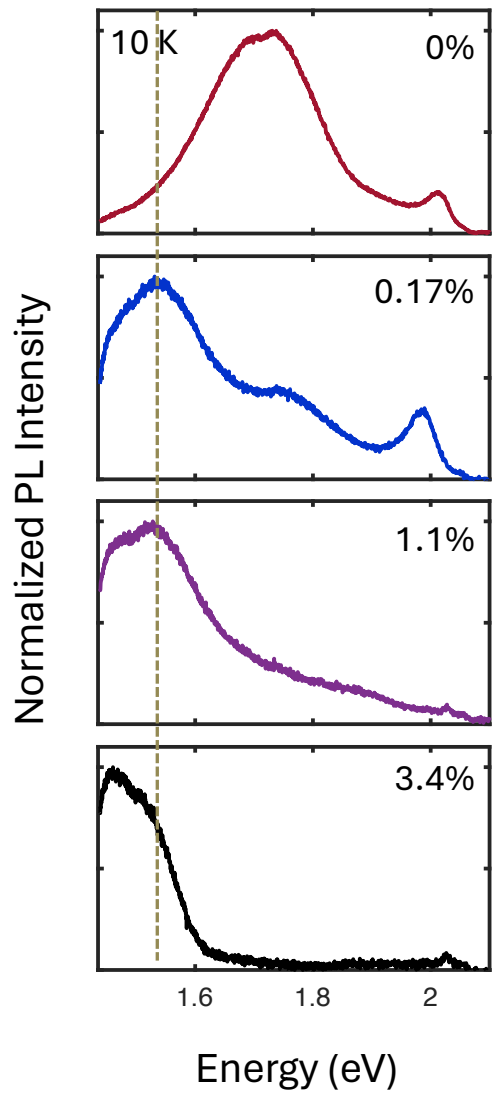

**Figure S5.** PL spectra of pristine, 0.17%, 1.1%, and 3.4% V-doped WS$_2$ measured at 10 K. The dotted line, centered around peak X$^V$, serve as visual guides indicating the redshift of peak X$^V$ with increasing vanadium concentration.

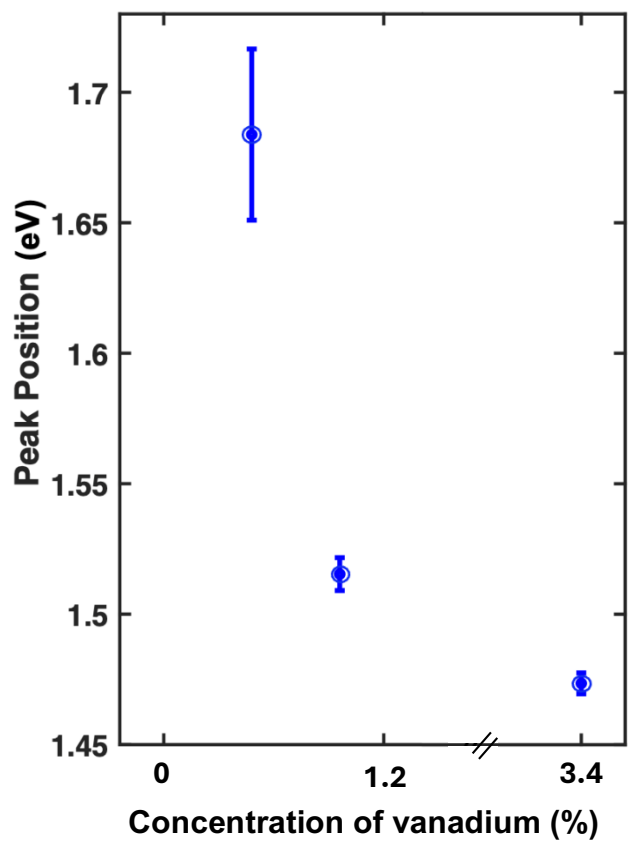

**Figure S6.** Graph showing the uncertainty limits of the peak $X^V$ in the V-doped $WS_2$ samples examined in this study.

**Note 2.** We note that the substrate can have a strong influence on the doping and electronic properties of TMDs such as $WS_2$.[3] Therefore, we eliminate potential influences from the substrate following the annealing process using the following procedure. The V-doped $WS_2$ sample was transferred onto hexagonal boron nitride (hBN) and then annealed in a sulfur environment. hBN is a chemically inert substrate known for its inability to introduce charge doping to the monolayer. The PL spectrum of this transferred sample does not exhibit defect-related excitonic emission after the S-annealing process. This control experiment demonstrates that the loss of the defect-related excitonic emission results from healing of sulfur vacancies rather than from charge doping introduced by substrate interactions. We note that the A-exciton peak in the S-annealed sample transferred to hBN appears broader in comparison to the samples that were not transferred. The origin of this effect is a topic of on-going investigation.

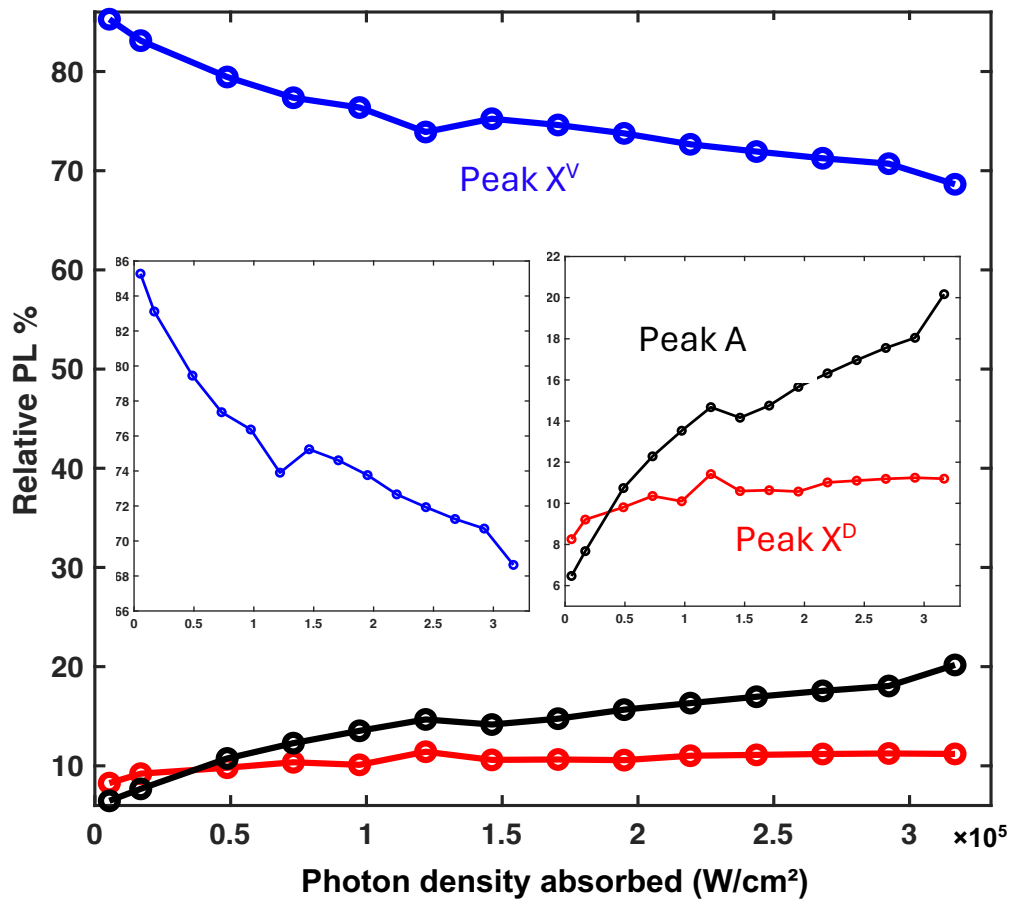

**Figure S7.** Relative integrated photoluminescence (RIPL) intensities of peak $X^D$, $X^V$, and A plotted against the absorbed photon density (W/cm²) in 3.4% vanadium (V)-doped $WS_2$ monolayers displaying saturation of the defect peak $X^V$ at moderate excitation, while peak $X^D$ and A grow, indicating the impact of vanadium-sulfur vacancy pairing on emission behavior at elevated excitation levels.

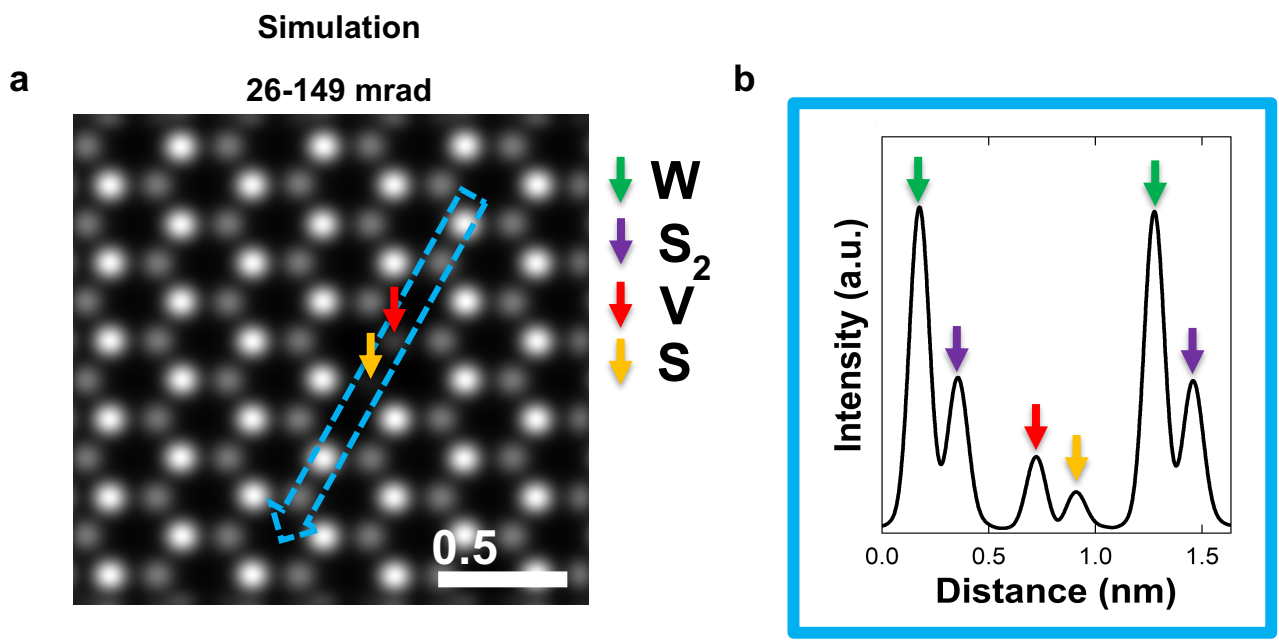

**Figure S8.** Multislice simulation of the expected atomic intensities for V-$S_v$ coupled sites: (a) ADF-STEM image of V-doped $WS_2$ and (b) an intensity line profile (marked in blue on the image in a) showing an $S_v$ site adjacent to a V dopant.

a

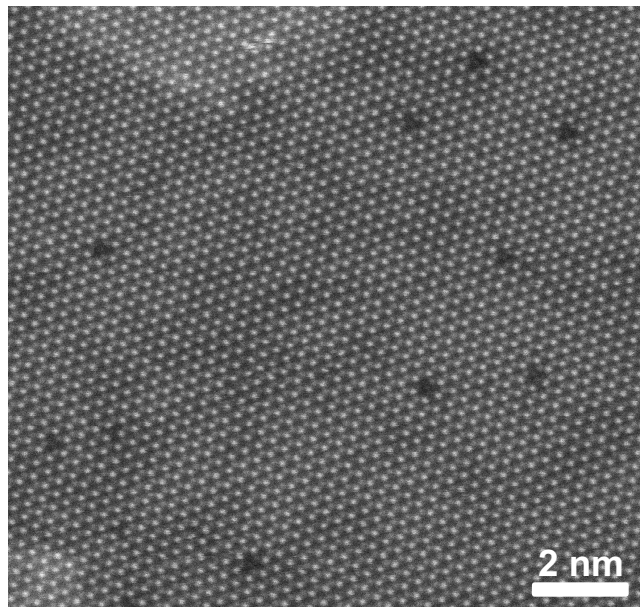

**Raw**

b

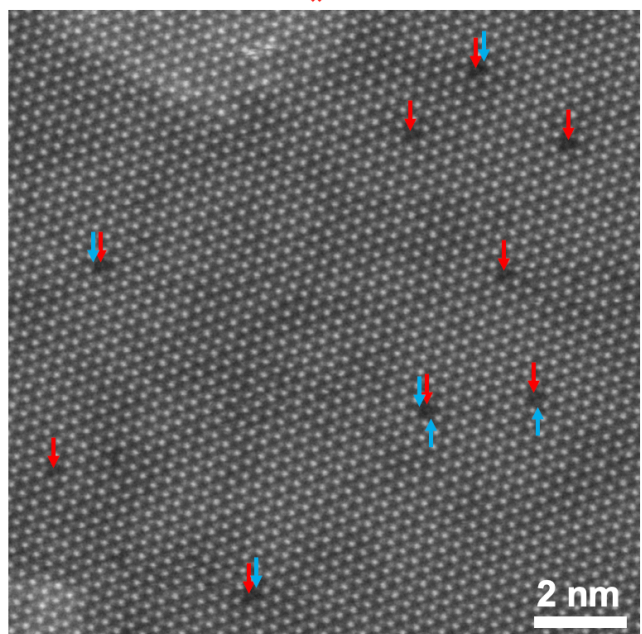

**Gaussian blurred (sigma = 2)**

**Figure S9.** ADF-STEM image of 0.3% V-doped $WS_2$ with $S_V$ sites (blue) adjacent to V dopants (red) marked: (a) raw image and (b) Gaussian blurred (sigma = 2).

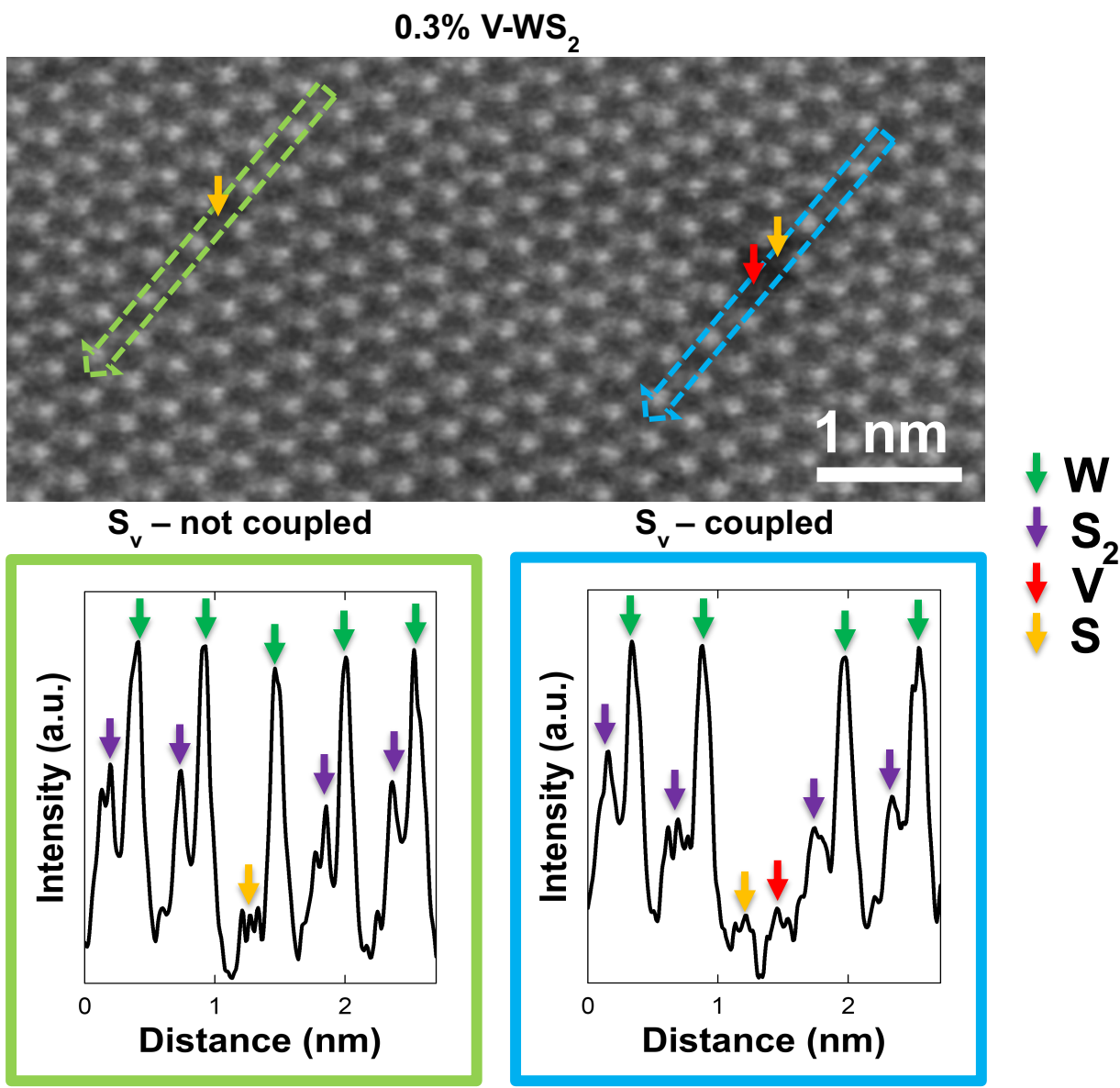

**Figure S10.** ADF-STEM image and intensity line profiles showing $S_V$ sites not coupled (green) and coupled (blue) to V dopants.

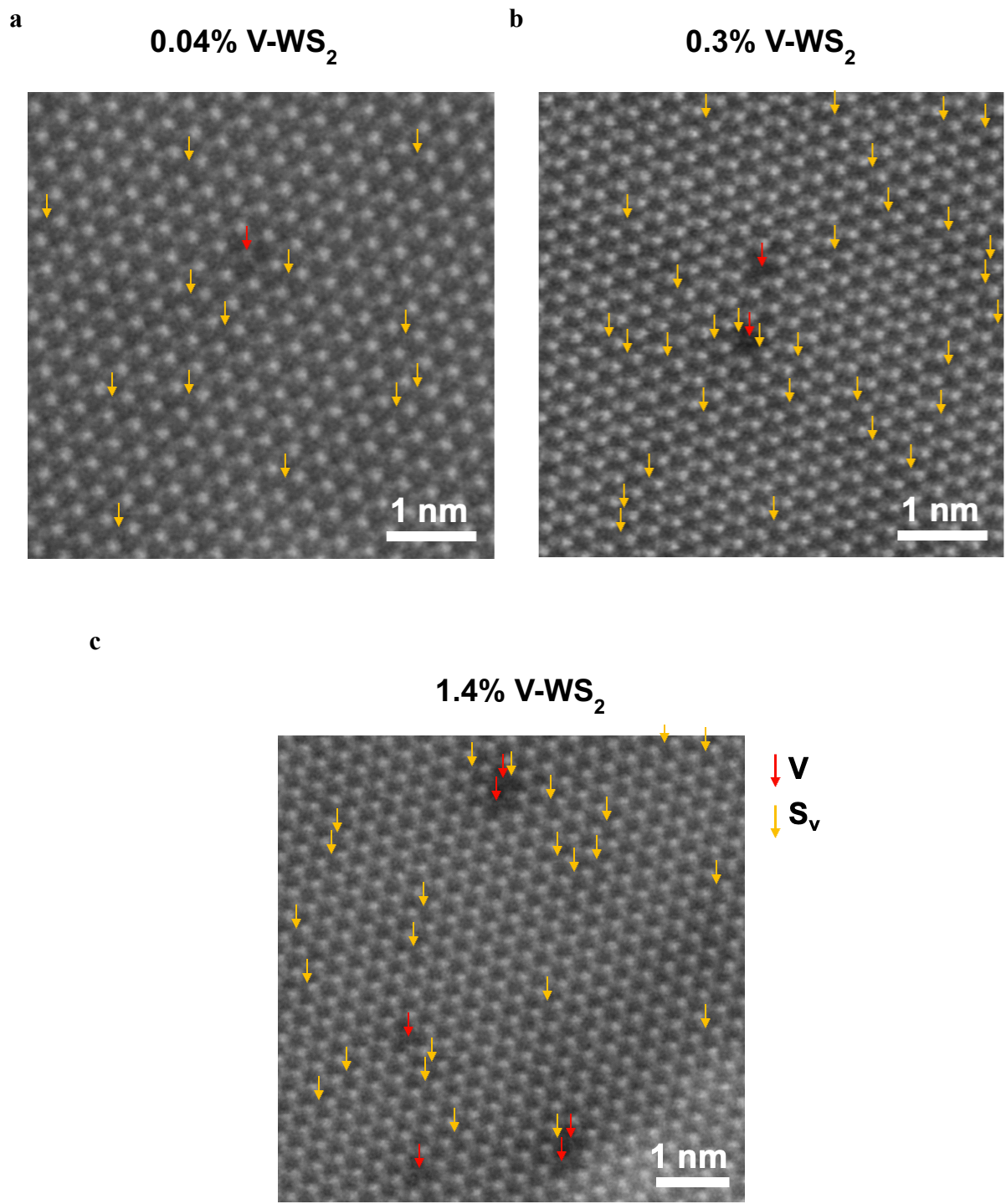

**Figure S11.** ADF-STEM images of WS$_2$ marked with V dopant (red) and S$_V$ (orange) sites, with V concentrations of (a) 0.04%, (b) 0.3%, and (c) 1.4%.

## V-S$_V$ Paired WS$_2$-1

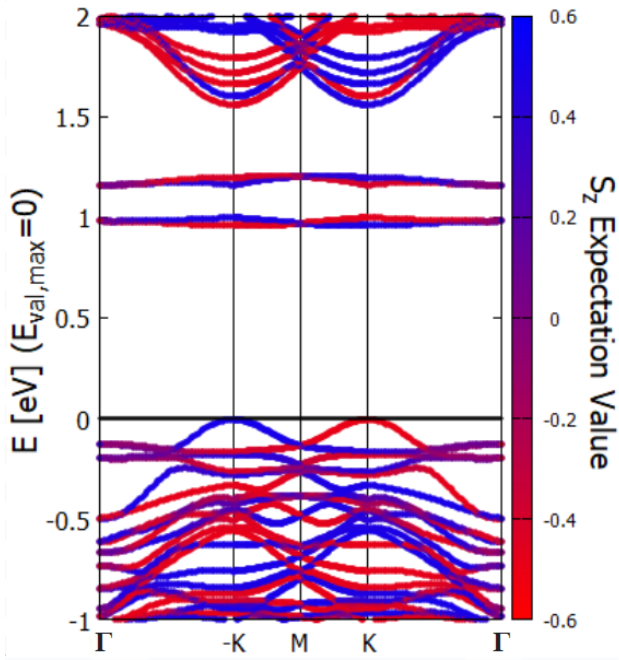
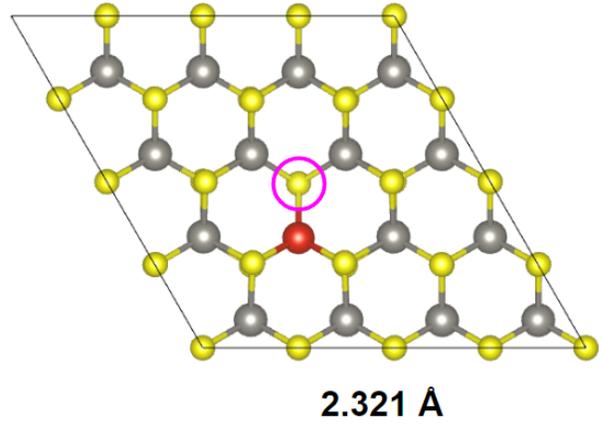

$E_{trans} \sim 1.56$ eV

2.321 Å

## V-S$_V$ Paired WS$_2$-2

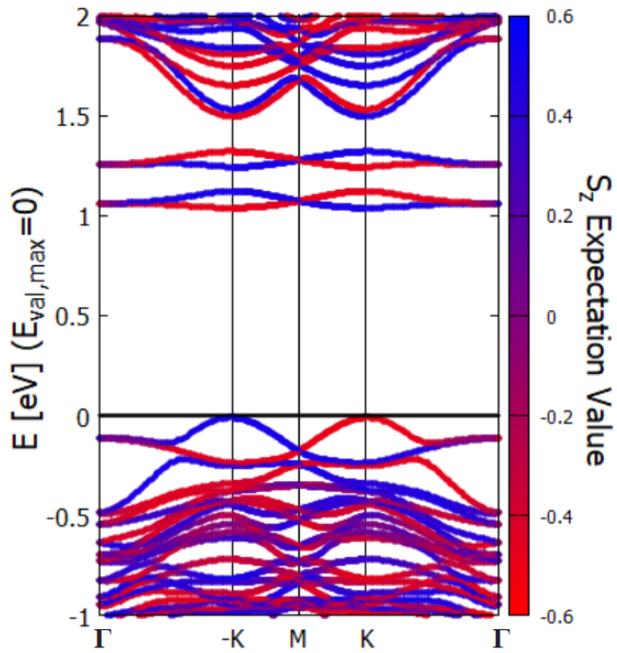
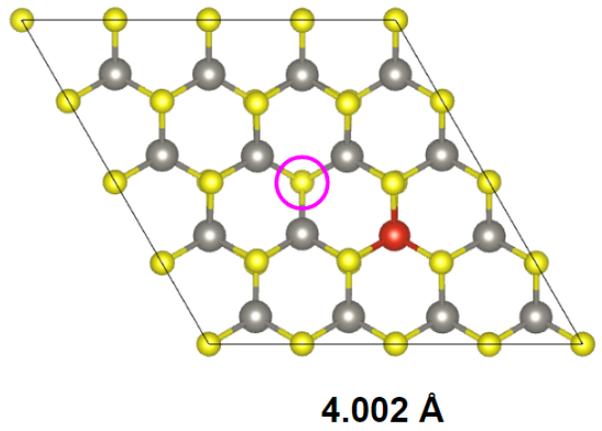

$E_{trans} = 1.8$ eV

4.002 Å

## V-S$_V$ Paired WS$_2$-3

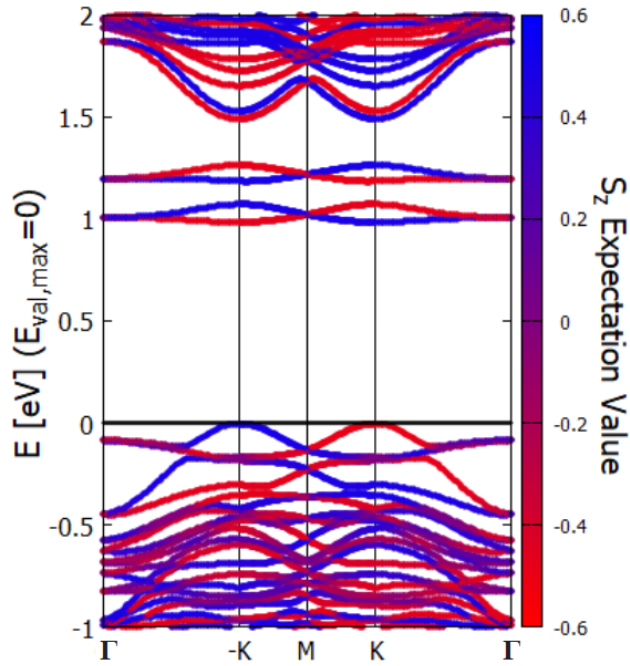

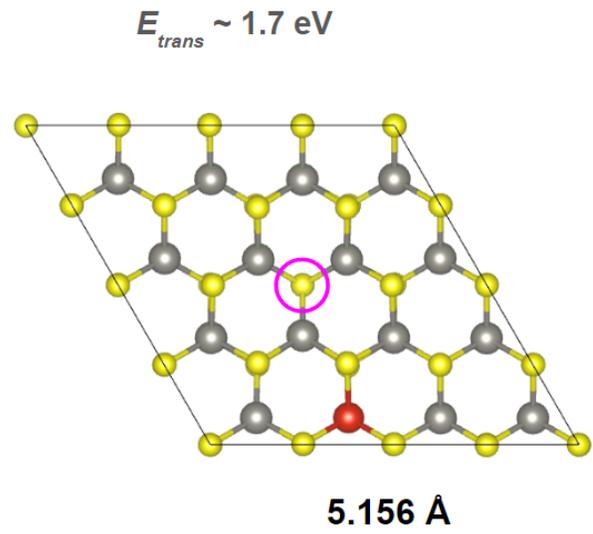

$E_{trans}$ ~ 1.7 eV

**5.156 Å**

## V-S$_V$ Paired WS$_2$-4

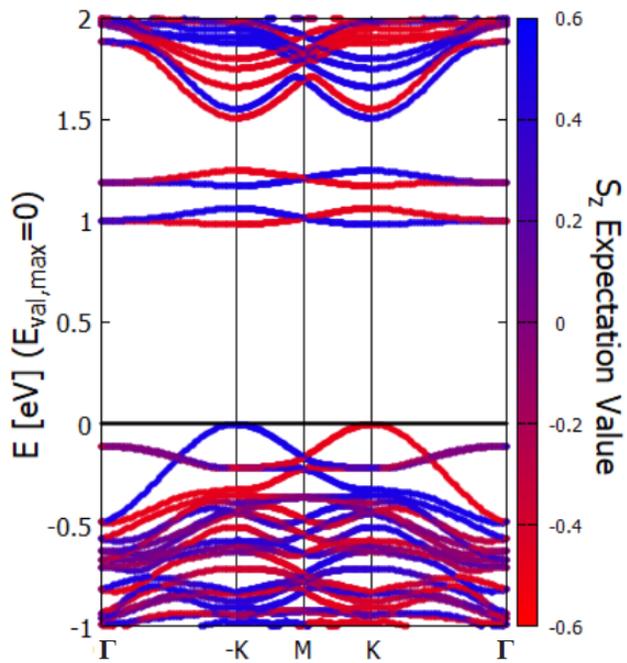

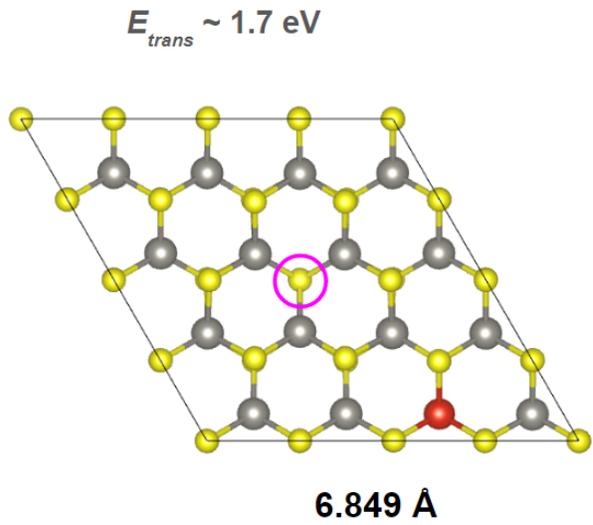

$E_{trans}$ ~ 1.7 eV

**6.849 Å**

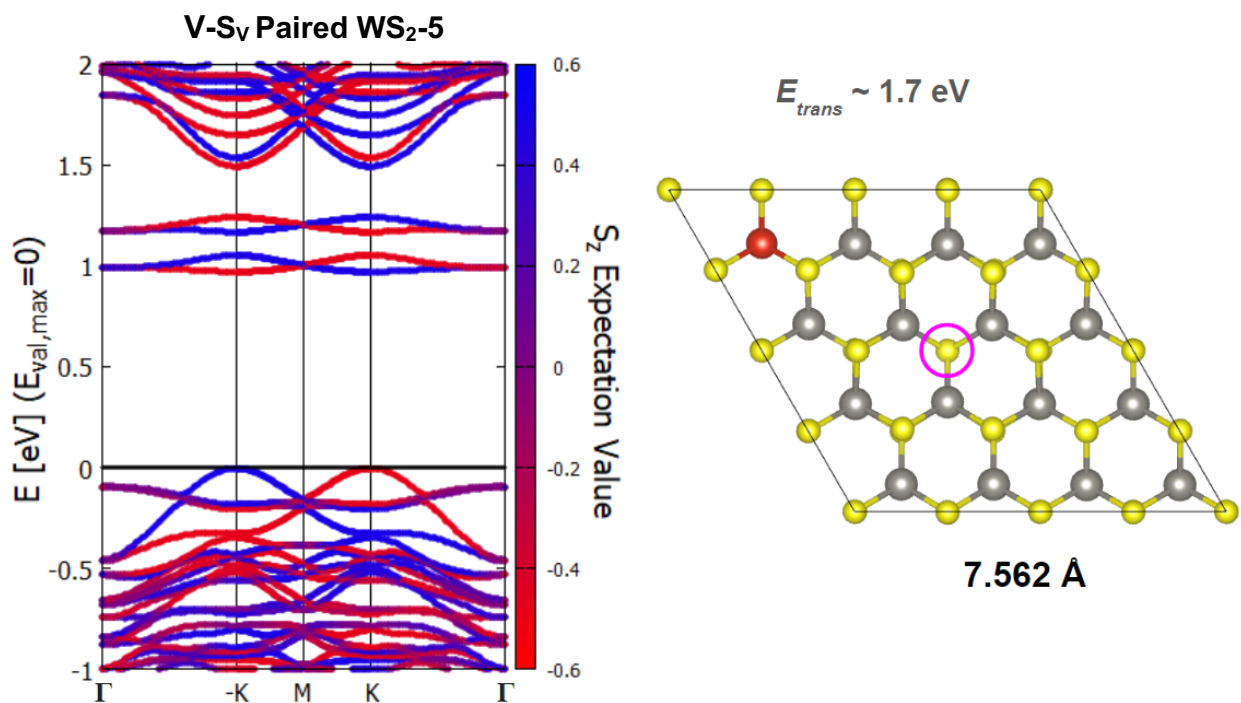

**Figure S12.** The band structures obtained for the V-S$_V$ defects using spin-orbit coupling where the corresponding structure is shown to the right with the V atom in red and the S$_V$ defect in pink. The $E_{trans}$ corresponds to the shifted transition energy and the bold distance is the distance between the V atom and the S$_V$ defects.

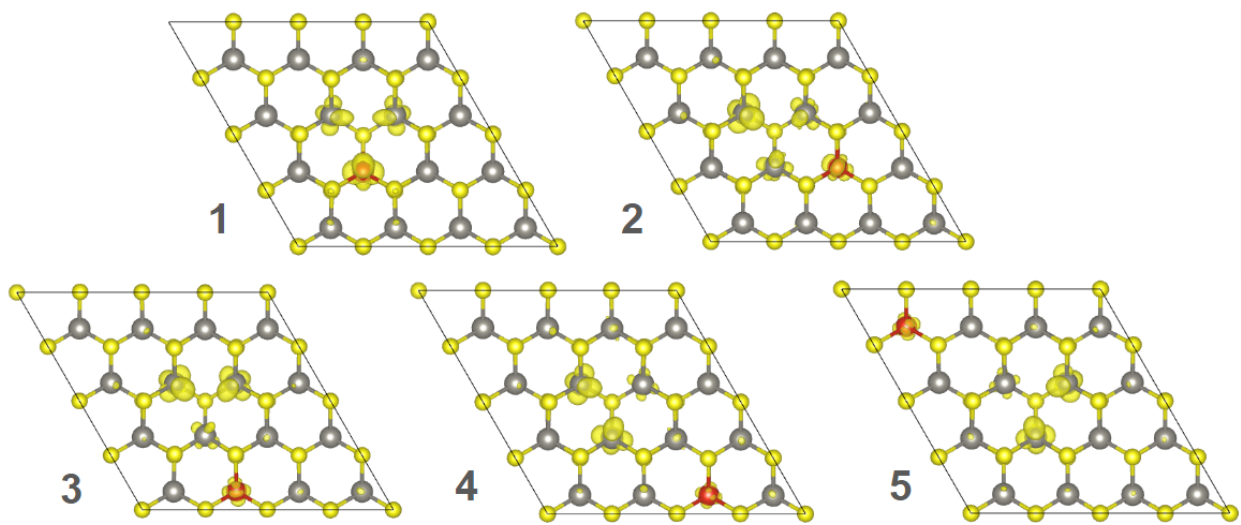

**Figure S13.** The wavefunction overlaps for the blue band transitions ($v_i$-$d_1$) with the same configurations shown in Fig. S12.

**Note 3**. For **Figure 5a**, the sulfur vacancy is maintained at a fixed position, while the vanadium dopant is adjusted to increase the distance between the V dopant and S vacancy defects. The total energy is determined for five configurations with the V dopant positions colored red, orange, green, blue, and purple for increasing distance between the defects. The same configurations shown in **Figure 5a** are used with the HSE calculated[4] band gaps shown in **Figure 5b,** and the splitting between the defect bands is obtained from HSE DOS calculations for **Figure 5c**. For **Figure 5d** and **S12**, the band structures in the spin-up (blue) and spin-down (red) states with the corresponding supercells are shown for the $\Gamma \rightarrow -K \rightarrow M \rightarrow K \rightarrow \Gamma$ path in the Brillouin zone with the transition energies calculated from the highest valence blue band to the lowest blue defect band at the K point. For **Figure S13**, the wavefunction is evaluated at the bands corresponding to the highest valence blue band to the lowest blue defect band at the K point, and then the wavefunction overlap (or product) is determined between those two wavefunctions. For convenience, we define the zero of the relative energy scale in **Figure 5a** for a vanadium atom adjacent (0.2 nm) to a monosulfur vacancy defect. Then, we computed the energy above this hull value for configurations in which the vanadium is located at nearby tungsten sites of increasing distance, up to 0.8 nm separation.